\documentclass[%
 reprint,
 amsmath,amssymb,
 aps,
]{revtex4-2}

\usepackage{graphicx}
\usepackage{dcolumn}
\usepackage{bm}
\usepackage{xcolor}
\usepackage{hyperref}
\usepackage{float}

\begin{document}

\preprint{APS/123-QED}

\title{Ion counting and temperature determination of Coulomb-crystallized laser-cooled ions in traps using convolutional neural networks}

\author{Yanning Yin}
\email{yanning.yin@unibas.ch}
\author{Stefan Willitsch}%
\affiliation{%
 Department of Chemistry, University of Basel, Klingelbergstrasse 80, 4056 Basel, Switzerland
}

\date{\today}

\begin{abstract}
Coulomb crystals—ordered structures of cold ions confined in ion traps—find applications in a variety of research fields. The number and temperature of the ions forming the Coulomb crystals are two key attributes of interest in many trapped-ion experiments. Here, we present a fast and accurate approach to determining these attributes from fluorescence images of the ions based on convolutional neural networks (CNNs). In this approach, we first generate a large number of images of Coulomb crystals with different ion numbers and temperatures using molecular-dynamics simulations and then train CNN models on these images to classify the desired attributes. The classification performance of several common pretrained CNN models was compared in example tasks. We find that for crystals with ion numbers in the range 100--299 and secular temperatures of 5--15~mK, the best-performing model can discern number variations on the level of one ion with an accuracy of 93\% and temperature variations by 1 mK with an accuracy of 92\%. Since the trained model can be directly integrated into experiments, \emph{in-situ} determination of these attributes can be realized in a non-invasive fashion, which has the potential to greatly facilitate the analysis and control of trapped ions in real time.
\end{abstract}

\maketitle


\section{\label{sec:intro} Introduction}

Coulomb crystals—ordered structures of cold ions confined in ion traps—are an unusual form of matter that has been employed in a variety of research fields including quantum information and computation \cite{bruzewicz2019trapped, monroe2013scaling, haffner2008quantum, wineland2009quantum}, atomic clocks \cite{ludlow2015optical}, fundamental physics studies \cite{safronova2018search} and cold and controlled chemistry\cite{willitsch2012coulomb}. Coulomb crystals can be generated by loading ions into radio-frequency (RF) or Penning traps \cite{major2005charged} and cooling their motion to millikelvin or even lower temperatures using laser cooling (for suitable atomic ions) or sympathetic cooling (for atomic species that cannot be laser cooled and molecular ions). Since the ions are confined in deep traps in ultrahigh-vacuum chambers where collisions with gas and other perturbations from the surroundings are suppressed, Coulomb crystals can exhibit remarkable stability with lifetimes ranging from minutes to days, allowing the observation, addressing, and manipulation of the ions localized in the trap down to the single-   particle level. 

For a Coulomb crystal, the number of ions ($N$) and temperature ($T$) are two important attributes that are of interest in many experiments. For example, in studies of cold collisions between ions and other species, these two parameters and their variations can provide important information about the collision or reaction process in terms of collision energy and reaction rate \cite{willitsch08b}. During the past years, different methods have been developed to determine $N$ and $T$ for ions in Coulomb crystals. While $N$ is often obtained by counting ions using fluorescence imaging \cite{Schmid2022number} or a time-of-flight mass spectrometer (TOF-MS) \cite{schneider2014laser, meyer2015ejection, schmid2017ion, roesch16a}, the determination of $T$, often referred to as ion thermometry, is less straightforward, and various methods have been developed that are applicable to different cases. For ions in Coulomb crystals with temperatures $T > 1$ mK, thermometry techniques include the analysis of Doppler-broadened line shapes\cite{Herrmann2009Frequency} and dark resonances \cite{Roßnagel2015fast}, spatial imaging of single ions and small ion chains \cite{norton2011millikelvin, knunz2012sub, rajagopal2016trapped} as well as of large crystals\cite{Prestage1991, Ostendorf2006, Zhang2007, Okada2010, Tong2010}. For lower temperatures with ions cooled close to their motional ground state in the trap, resolved-sideband techniques \cite{Chen2020Efficient, Feng2020Efficient, Sawyer2012Spectroscopy, DOnofrio2021}, applicable mainly to ion chains and small crystals, as well as extensions of this technique \cite{Vybornyi2023Sideband} applicable to large crystals, have been demonstrated in the past. For higher temperatures, Doppler-recooling \cite{Wesenberg2007Fluorescence, Epstein2007Simplified, Zipkes2010A, Sikorsky2017Doppler}  and time-of-flight experiments \cite{notzold2020thermometry} have been used. 

For Doppler-cooled Coulomb crystals, many of the ion-counting and thermometry methods mentioned above rely on interference with the trapped ions, either by destroying the crystals, such as in the time-of-flight method, where the ions are ejected from the trap, or potentially changing the ions' attributes, as in Doppler-broadened line-shape analysis, where the cooling-laser frequency is scanned and thus the ion temperature changes. In the image-analysis methods, images of large crystals taken during experiments are compared visually to images generated by molecular-dynamics (MD) simulations to infer $T$ and $N$ for the ions\cite{Prestage1991, Ostendorf2006, Zhang2007, Okada2010, Tong2010}. In the past, this visual comparison had to be performed \emph{a posteriori} in a time-consuming, iterative simulation procedure and was not implemented in real-time, which prevented a rapid and \emph{in-situ} characterization of the Coulomb crystals during experiments.

Here, we present a fast, nonperturbative, and accurate method to determine these two attributes based on convolutional neural networks (CNNs). In this approach, we first generated a large number of images of crystals with different $N$ and $T$ values using MD simulations and then trained CNN models on these images to classify these attributes. By comparing the classification results of several popular pretrained CNN models, we identified the best models for determining $T$ and $N$, respectively. 
The trained model can be applied to the classification of experimental images and can also be integrated directly into experiments, making real-time and \emph{in-situ} determination of crystals' characteristics feasible, which offers the potential to greatly enhance the analysis and control of trapped ions in experiments. 

The remainder of this paper is structured as follows. Sec. \ref{sec:method} explains the methods for generating a sufficient number of simulated images for training CNN models, and it presents the details of the selection of CNN models as well as the training and validation process. Sec. \ref{sec:results} provides some examples of the application of  CNN models to image-classification tasks, compares the performances of several CNN models, and discusses the application of the trained models to classifying experimental results. Sec. \ref{sec:conclusion} summarizes the main results and suggests potential avenues for further improvements.

\section{\label{sec:method} Methods}

\subsection{Image generation}

\begin{figure}[t]
    \includegraphics[width=\linewidth]{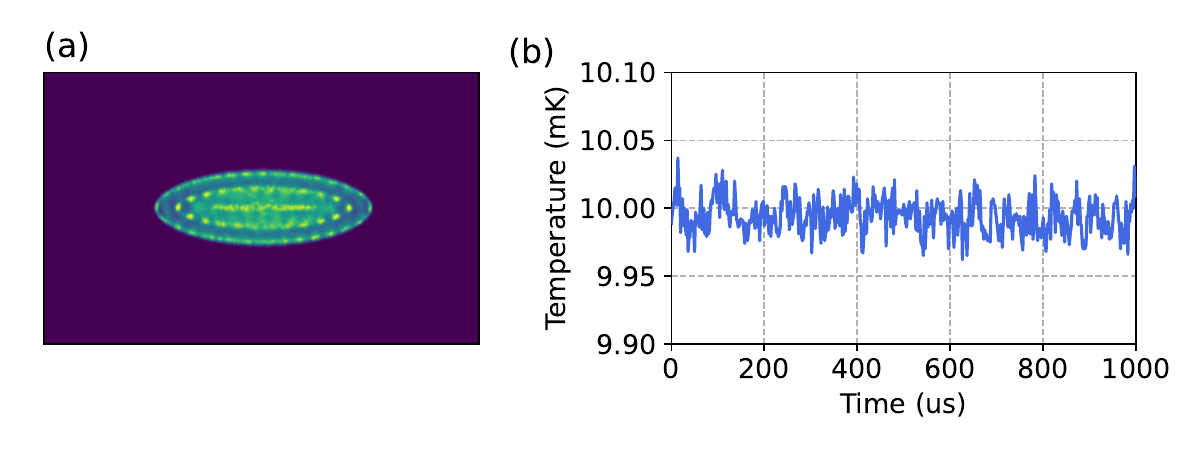}
    \caption{\label{fig:fig1}(a) An example false-color image of a Coulomb crystal with 200 ions and a secular temperature of 10 mK. (b) Fluctuation of the temperature after stabilization extracted from the simulation generating the crystal in (a). }
\end{figure}

\begin{figure*}[t]
    \includegraphics[width=\textwidth]{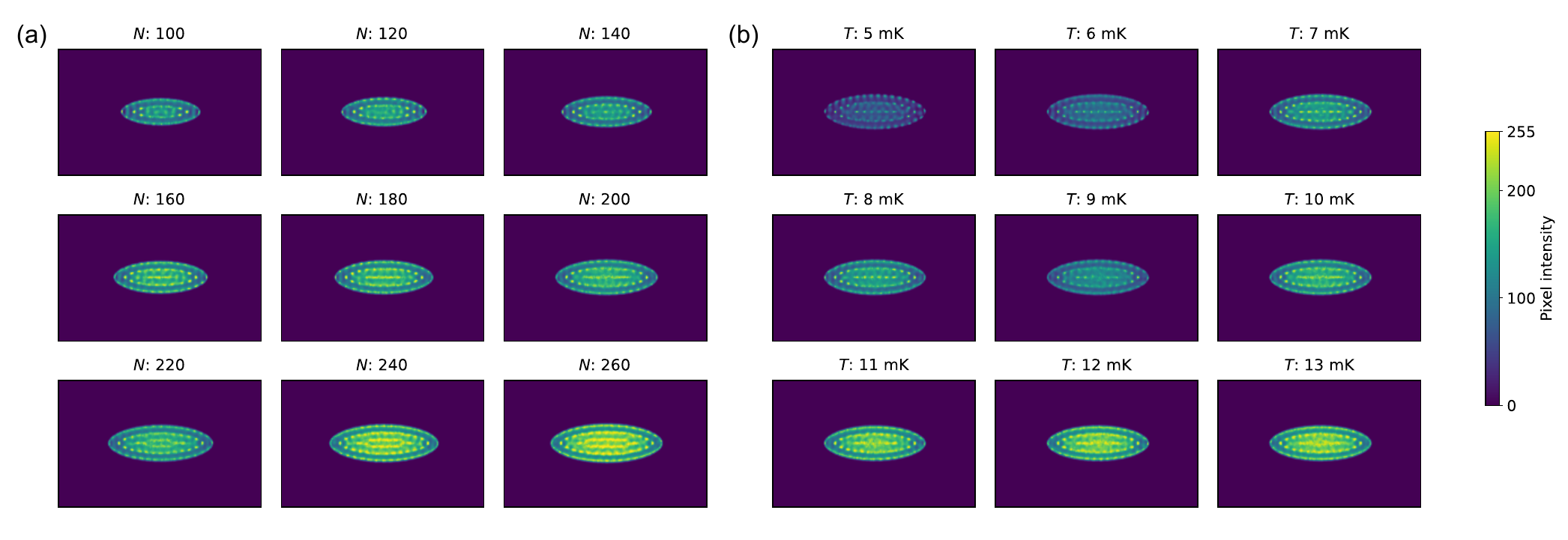}
    \caption{\label{fig:fig2}(a) Example false-color images of Coulomb crystals with different numbers of ions for a fixed secular temperature of $T=10$ mK. (b) Example images of Coulomb crystals with different secular temperatures for a fixed number of ions $N=200$. The color bar applies across all figures showing crystal images in the article.}
\end{figure*}

For training a CNN model, both the quality and quantity of images input into the model are crucial for its ultimate performance in terms of accuracy and reliability. Although experimental images of Coulomb crystals taken by cameras would be ideal sources for this task, the collection of such a vast amount of data would be painstaking. To our knowledge, there is currently no systematic collection with a sufficient number of images covering a wide range of well-determined values for the salient attributes available. Since image comparison between simulations and experiments is routinely employed as a method for determining $N$ and $T$, generating a large number of images with varying attributes by simulation was pursued as an efficient way to obtain suitable images for training the CNN model, which was then transferred to the classification of experimental images.

To generate images of Coulomb crystals with well-defined numbers and temperatures of ions, we adapted the MD simulation method described in Ref.~\cite{rouse15a}. Here, the trajectories of $N$ ions inside an RF ion trap were simulated by taking into account the different forces acting on the trapped ions. The force $\boldsymbol{F}_{\rm{i}}$ exerted on an ion $i$ can be written as

\begin{equation}
    \boldsymbol{F}_{\rm{i}} = -\nabla U(\boldsymbol{r}, t)+\boldsymbol{F}_{\rm{Coulomb}}+\boldsymbol{F}_{\rm{cooling}}+\boldsymbol{F}_{\rm{bg}},
\end{equation}

where $U(\boldsymbol{r}, t)$ is the time-dependent trapping potential, 
$\boldsymbol{F}_{\rm{Coulomb}}$ is the Coulomb interaction between ions, $\boldsymbol{F}_{\rm{cooling}}$ is the cooling force from photon scattering, and $\boldsymbol{F}_{\rm{bg}}$ is an effective heating force assumed to be dominated by collisions with background gas, which also effectively includes other heating mechanisms such as RF heating \cite{rouse15a}.

Generally, the trap potential $U(\boldsymbol{r}, t)$ is determined by the trap geometry and experimental parameters (DC voltages and the RF amplitude and frequency applied to different electrodes). The specific implementation of the trap varies from one experiment to another. In the present work, we used the harmonically approximated potential of a segmented RF trap, under the experimental conditions described in detail in Ref.~\cite{mangeng2023}. The potential takes the form

\begin{equation}\label{eq:eq2}
    U(\boldsymbol{r}, t) = \frac{m\omega^2}{8}\sum_i(a_i-2q_i\cos{\Omega{t}})r_{i}^2,
\end{equation}

where $m$ is the mass of the ion, $\omega$ is the applied RF frequency, and $r_i$ is the ion position coordinates. 
Here, $\Omega$ was set to $2\pi\times10 $ MHz, and the Mathieu parameters $a_i$ and $q_i$ \cite{major2005charged} were derived from numerical modeling: $a_x=-4.62\times10^{-4}$, $a_y=-4.10\times10^{-4}$, $a_z= 8.83\times10^{-4}$, $q_x= 8.73\times10^{-2}$, $q_y=-9.15\times10^{-2}$, and $q_z= 0$ thus mimicking realistic experimental values.

MD simulations were carried out using the OpenMM framework \cite{openmm}, a high-performance, open-source toolkit supporting both built-in force fields, such as Coulomb interactions, and custom forces. This framework also enables the computations to be sped up using graphics-processing-unit (GPU) acceleration. The cooling force $\boldsymbol{F}_{\rm{cooling}}$ was modeled by momentum kicks of magnitude $\hbar{k}$ due to photon absorption or emission applied to the ions as they undergo transitions between ground and excited states during laser cooling \cite{rouse15a} . 

While the number of ions $N$ can simply be specified as an input parameter for the MD simulations, the temperature of the ions $T$ is less straightforward to control, as it depends on the dynamics of the ion ensemble in the presence of several competing interaction terms. One common way to regulate $T$ is to repeatedly apply a velocity kick with a random direction to the ions, which represents an overall effective heating effect for various heating processes \cite{Zhang2007}. However, as examined in Ref. \cite{rouse15a}, adding a velocity kick that essentially resembles an elastic collision of ions with other species (or virtual particles \cite{Okada2010}) can cause the ions to sample a broad range of temperatures, especially for infrequent kicks with a large momentum transfer. According to our tests, even adding continuous, weak velocity kicks can cause two severe problems for generating images with a well-defined $T$. First, the stability of $T$ during the simulation can easily be worse than $\pm$1~mK, which would lead to images that are difficult for the CNN model to classify with an accuracy on that level. Second, there is not a universal velocity-kick frequency and magnitude to generate images with a certain $T$ if their $N$ values are different, which means that a tedious trial-and-error procedure is needed to tune the kick parameters to obtain a desired $T$ once an image with a different $N$ is desired.

To address this issue, we introduced a direct stabilization of $T$ based on feedback control. A target temperature of ions $T_{\rm{target}}$ was set at the beginning of the simulation. The temperature of the ion ensemble $T_{\rm{period}}$ was calculated after each RF oscillation by averaging the kinetic energy of $N$ ions according to \cite{Zhang2007}:

\begin{equation}
    T_{\rm{period}} = \frac{m}{3Nk_B}\sum_i^N(\overline{v}_{ix}^2 + \overline{v}_{iy}^2 + \overline{v}_{iz}^2)
\end{equation}

where $\overline{v}_{ix}$, $\overline{v}_{iy}$, and $\overline{v}_{iz}$ are the velocity components of the $i$th ion averaged over one RF period. In this way, the contribution of the micromotion \cite{major2005charged} to the ion velocities was removed from the averaging, resulting in a temperature that purely reflects the \emph{secular}, i.e., thermal, motion of the ions in the trap. The difference between $T_{\rm{period}}$ and $T_{\rm{target}}$ was then employed as an error signal $e(t)$ to modify the velocities of the ions based on a proportional–integral–derivative (PID) controller. The modified velocity $u_{i}$ of the $i$th ion was related to its previous velocity $v_{i}$ by

\begin{equation}
    u_{i}^{2}= v_{i}^2 + K_Pe(t) + K_I\int e(t) \mathrm{d}t + K_D \frac{\mathrm{d}e(t)}{\mathrm{d}t}
\end{equation}

where $K_P$, $K_I$, and $K_D$ are the coefficients for the proportional (P), integral (I), and derivative (D) terms in the controller. In practice, we found that using only the P term was sufficient for stabilization.  The optimized parameters used in our simulations are $K_P=0.65$ and $K_I=K_D=0$. We then defined an average of $T_{\rm{period}}$ over 100 RF periods as the secular temperature of the ion ensemble $T$.

After $T$ was stabilized, the position of each ion after each time step (1 ns) was recorded and binned into a three-dimensional (3D) histogram of positions collected over a duration of 1~ms. From the 3D histogram, two-dimensional slices were extracted along the viewing direction of a camera. For each slice, a Gaussian blur was applied, the intensity of which was varied based on the slice's distance from the central layer to simulate the effect of the finite focal depth of the microscope attached to the camera. All slices were then combined, and the resulting image was normalized to its maximum intensity before being saved as the final simulated image. As an example, Fig. \ref{fig:fig1} (a) shows an image of a Coulomb crystal with $T$=10 mK and $N$=200 thus generated, and Fig.\ref{fig:fig1}(b) shows the corresponding time-dependent temperature after stabilization. It can be seen that the fluctuation of $T$ is within $\pm$0.04 mK over 1~ms due to the PID stabilization. This amounts to $\pm$4\% relative uncertainty with respect to 1~mK, which is sufficiently precise to serve as training data for a CNN model. In comparison to other approaches to simulating crystals relying on manual parameter tuning, our PID-based method offers automated and more precise temperature control, enabling efficient generation of high-quality training data, which is essential for training robust models.

In this way, a large number of simulated images with varying $N$ and $T$ were generated to train the models. As an example, Fig. \ref{fig:fig2}(a) shows a small collection of images of Coulomb crystals generated with different numbers of ions at a secular temperature of 10 mK, and Fig. \ref{fig:fig2}(b) displays crystals with 200 ions at different secular temperatures.

\subsection{CNN models}

To classify the images of Coulomb crystals in terms of $N$ and $T$, we employed one of the most widespread deep-learning methods for image classification:  CNNs \cite{lecun2015deep, krizhevsky2012imagenet}. CNNs can automatically and adaptively learn spatial hierarchies of features from input images by applying convolutional and pooling layers, and they have proven to be especially effective in image recognition \cite{krizhevsky2012imagenet, simonyan2014very, he2016deep, tan2019mnasnet}.

As a common practice when using CNNs, we adopted transfer learning \cite{yosinski2014transferable, oquab2014learning, donahue2014decaf} for the present image-classification task. This technique reuses a pretrained model from other tasks as the starting point for a new task, rather than starting from scratch. During transfer learning, the initial layers of the CNN models can capture generic features while the later layers are fine-tuned to the new task. This approach leverages the knowledge gained from the initial training to accelerate and enhance the new learning process, thus leading to faster convergence and better performance.

Among a large number of pretrained CNN models that are available for transfer learning, we selected a few widely used variants—AlexNet \cite{krizhevsky2012imagenet}, ResNet18 \cite{he2016deep}, VGG16 \cite{simonyan2014very}, and MnasNet \cite{tan2019mnasnet}—for our task in consideration of a balance between their performance and computational efficiency. The definitions of these models, including their architectures and the pretrained weights, are accessible in the \texttt{torchvision.models} module of PyTorch \cite{paszke2019pytorch}, an open-source machine-learning framework that was used in the present work. For example, we can specify the model architecture and load the pretrained weights by calling

\begin{verbatim}
    from torchvision import models  
    net = models.resnet18(pretrained=True)
\end{verbatim}

This loads the ResNet18 model with weights pretrained on the ImageNet dataset \cite{deng2009imagenet}, a large-scale benchmark dataset containing over one million labeled images across 1000 object categories.

\subsection{\label{subsec:training} Training and validation}

To use the generated images for training a CNN model, the range of $N$ (or $T$) to be classified by the model needs to be specified. For example, a given range of $N$=100--299 with a step of 1 will train the model to classify the number of ions in the crystals to differences within one ion count. The specified images are resized to 224$\times$224 pixels as required by the pretrained CNN models mentioned above. In addition, the original grayscale images can be converted to Red–Green–Blue (RGB) images using the \texttt{Image.convert(`RGB')} method provided by the Python Imaging Library \cite{pillow}, which replicates the grayscale channel across the three RGB
channels. This transformation does not introduce new information but ensures compatibility with CNNs pretrained on RGB images. Interestingly, in the present classification, it turned out that grayscale images outperformed RGB images for classifying $N$, whereas RGB images gave better accuracy for classifying $T$. This difference could be explained by noting that for classifying $N$, the shapes and sizes of crystals are more important features than colors, whereas for classifying $T$, the additional two color channels of RGB images could offer additional dimensionality for feature extraction, enabling the model to capture more complex patterns across channels. 

As we will discuss below, $N$ could be reliably determined by the present models with high accuracy for a much larger number of classes than $T$.
Although it is possible to train a model that can simultaneously classify both $N$ and $T$, known as multilabel classification, we found that the classification accuracy was significantly lower compared to training models for $N$ or $T$ separately. Therefore, in the remainder of the article, we will focus on the training of individual models.

As an example of classifying $N$=100--299 (in steps of 1 ion), we prepared a total of 7200 images, i.e., 36 images for each class of $N$, and we divided them randomly into training and validation datasets in the ratio 4:1. A batch size of 32, chosen based on our GPU capacity, was used for both training and validation. All the CNN models mentioned above can be employed either with or without their pretrained weights (PTWs); the latter means that only the architecture of the model is used. A comparison of the classification performances of different models with or without using PTWs will be given in the next section, with the help of specific image-classification examples.

After loading the model, its first convolutional layer was modified to accept single-channel input if the images were grayscale, as all the models by default accept three channels for RGB images. Moreover, the last layer of the model was also modified according to the number of classes. In our example, the last layer was adapted to 200 classes for classifying $N$=100--299 with a step of 1. We used the cross-entropy loss function \cite{krizhevsky2012imagenet, Goodfellow-et-al-2016} to measure the dissimilarity between the predicted probability distribution and the true distribution of the target classes, and the stochastic-gradient-descent (SGD) optimizer with a learning rate of 0.0006 and momentum of 0.9 to update the CNN parameters during the training loop \cite{Goodfellow-et-al-2016}. 

To accelerate the training and validation process, we employed a GPU (NVIDIA GeForce RTX 2080 Ti) for tensor operations in PyTorch, which could typically reduce the whole training and validation process of 200 epochs to within two hours.

\section{\label{sec:results} Results and Discussion}

\subsection{\label{subsec:examples}Image-classification examples}

\begin{figure}[h]
    \includegraphics[width=\linewidth]{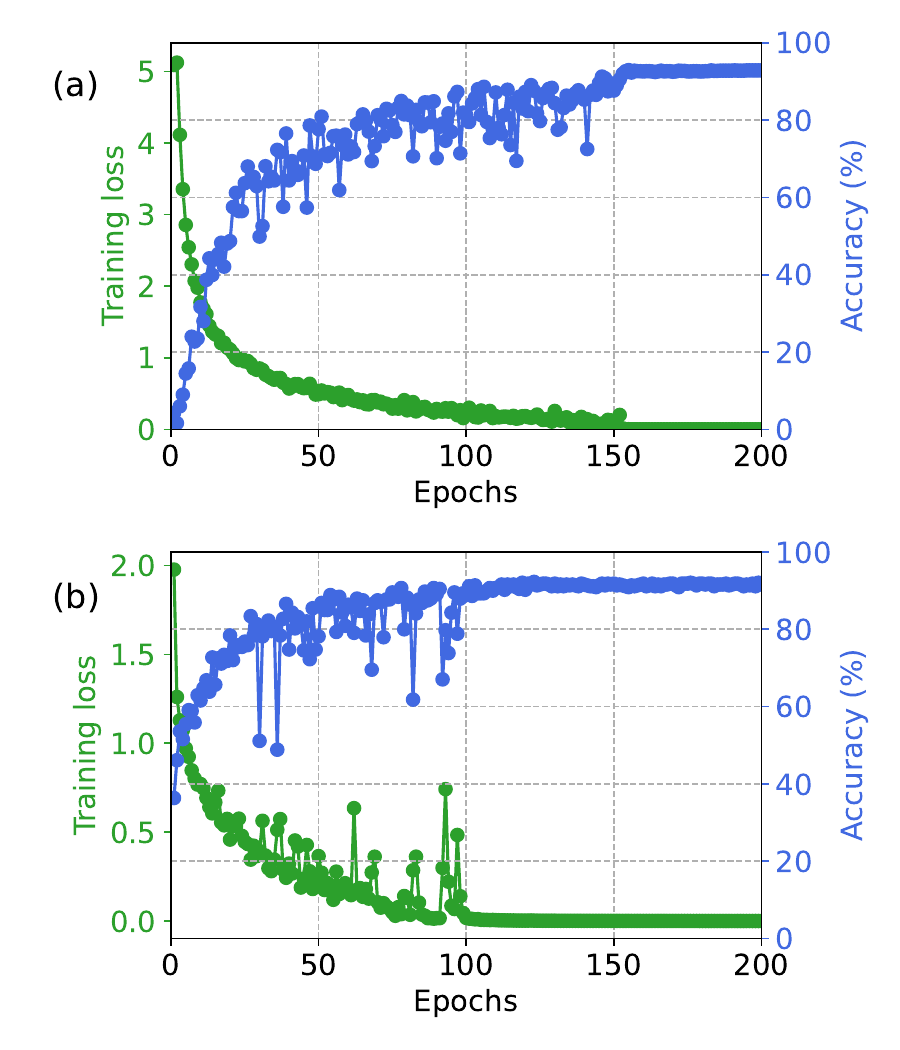}
    \caption{\label{fig:fig3} Training loss (in green) and validation accuracy (in blue) for classifying (a) $N$ in the range of 100--299 ions in steps of 1 ion using the CNN model ResNet18, (b) $T$ in the range of 5--15 mK in steps of 1~mK using AlexNet.}
\end{figure}

To demonstrate the performance of the CNN models, we present examples of image classification with $N$ in the range 100–299 and $T$ between 5 and 15~mK. This range is representative of typical ion-trap experiments involving medium-sized Coulomb crystals under Doppler cooling. Focusing on these conditions allows us to benchmark the method against realistic experimental settings and establishes a foundation for extention to broader regimes.  In principle, the approach is scalable to other $N$ and $T$ ranges, as it primarily requires training models on newly simulated (or experimentally collected) images under the relevant conditions.

For the example of classifying images with $N$=100–299 in steps of 1 ion mentioned above, the training loss and validation accuracy during training are plotted as a function of training epoch in Fig. \ref{fig:fig3}(a), for which the CNN model ResNet18 with its PTWs was employed. The training loss refers to the average cross-entropy loss across all batches in the epoch \cite{krizhevsky2012imagenet, Goodfellow-et-al-2016}, while the validation accuracy refers to the percentage of correctly predicted classes among all predictions made on the validation dataset. During the training loop, the training loss first dropped sharply for the first tens of epochs and then decreased steadily before it stabilized to $\leqslant0.001$  after about 157 epochs. The validation accuracy showed an opposite trend, increasing from less than 5\% to a final stabilized value of 93\% after 155 epochs, thus becoming increasingly capable of making correct predictions. This clearly shows that the model's performance improved significantly during training and could eventually distinguish validation images differing by a single ion with an accuracy better than 90\%.

As an example for classifying images of Coulomb crystals with $T=$5--15~mK in steps of ~1~mK, we loaded a total of 2200 images for training, i.e., 200 images are assigned to each $T$ class. The pretrained model AlexNet was used in this example. Apart from the fact that the images were converted to RGB, as explained in Sec. \ref{subsec:training}, the same procedures and hyperparameters (such as batch size, learning rate, number of epochs, etc.) as in the case for classifying $N$ were taken for the training and validation processes. The training loss and validation accuracy during training are shown in Fig. \ref{fig:fig3}(b). Both the loss and accuracy converged after $\approx$120 epochs, earlier than the case for $N$ in Fig. \ref{fig:fig3}(a). The final accuracy of 92\% highlights the model's ability to differentiate images with a temperature difference of 1~mK.

\begin{figure}[t]
    \includegraphics[width=\linewidth]{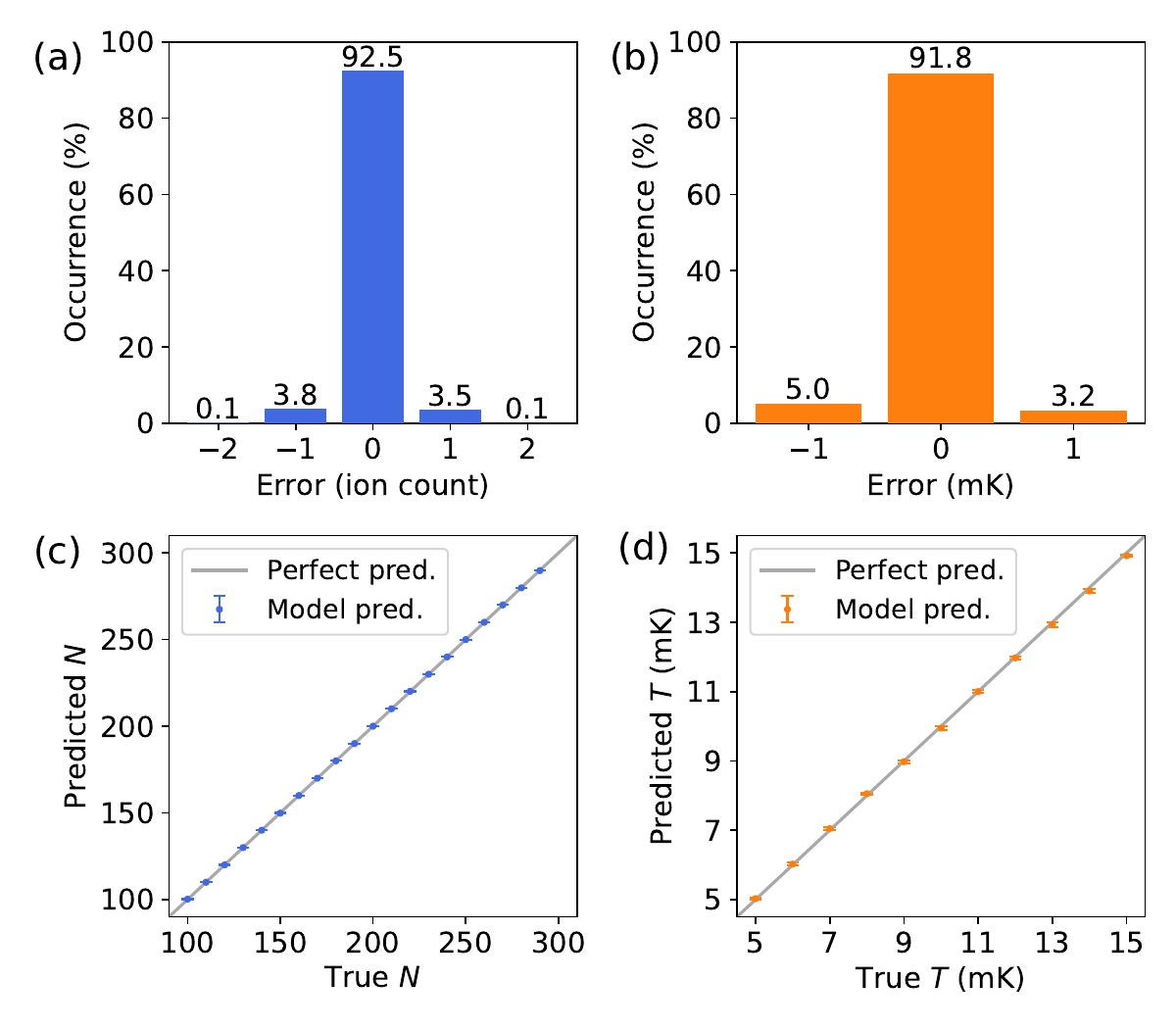}
    \caption{\label{fig:error_distribution}(a) Histogram of errors when using the trained model ResNet18 to classify 1440 validation images with respect to $N$ in the range 100--299 ions. (b) Histogram of errors when using the trained model AlexNet to classify 440 validation images with respect to $T$ in the range 5--15 mK. (c) Predicted $N$ by the model vs the true $N$ of the input images. Only a proportion of the datapoints in the range $N$=100--299 ions are shown for clarity. (d) Predicted $T$ vs true $T$.}
\end{figure}

To evaluate the distribution of errors, defined as the difference between the model's predicted class label and the true class label, for the trained models, histograms of errors when applying the models to validation images for the two examples are plotted in Figs. \ref{fig:error_distribution}(a) and \ref{fig:error_distribution}(b). Both histograms indicate a narrow error distribution, with variances of 0.0806 ion counts and 0.0815 mK [Figs. \ref{fig:error_distribution}(a) and \ref{fig:error_distribution}(b), respectively]. The histogram in Fig. \ref{fig:error_distribution}(a) is slightly more symmetric than that in Fig. \ref{fig:error_distribution}(b). This suggests that there are no systematic biases when applying the model to classify $N$ for the example, whereas there is a small bias toward lower temperatures in classifying $T$, although the errors are centered closely on zero (with means of $-$0.004 and $-$0.018, respectively). Figures \ref{fig:error_distribution}(c) and \ref{fig:error_distribution}(d) show the models' predictions for $N$ and $T$ versus the true labels of the input images, indicating both high accuracy and high precision.

Note that in the above examples, we let the training loop complete a total of 200 epochs for the sake of illustration. However, to avoid possible overfitting after the convergence—whereby the model becomes too specialized in the training data but fails to perform well on new, unseen data \cite{caruana2000overfitting, prechelt2002early, hawkins2004problem, yao2007early}—in practice, we terminated the training process earlier at the point when the validation accuracy ceased to improve for ten consecutive epochs. The model, including its architecture and updated weights, was saved at the epoch in which the highest accuracy was achieved.

\subsection{\label{subsec: model_comparison} CNN model comparison}

\begin{table}[b]
\caption{\label{tab:tab2} Validation accuracies after convergence of different CNN models for classifying $T$ in the range of 5--25~mK (in steps of 2 mK).}
\begin{ruledtabular}
\begin{tabular}{ccccc}
Model & AlexNet&ResNet18&MnasNet&VGG16\\
\hline
Accuracy & 94\%&98\%&97\%&95\% \\
\end{tabular}
\end{ruledtabular}
\end{table}

\begin{figure}[t]
    \includegraphics[width=\linewidth]{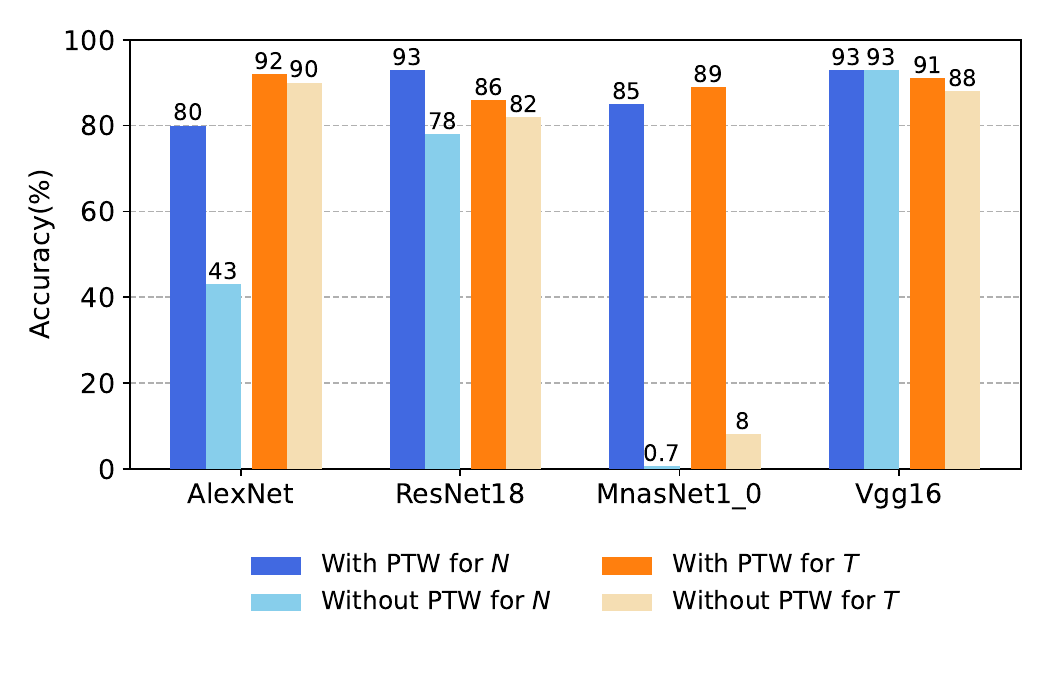}
    \caption{\label{fig:accuracies} Validation accuracies after convergence for different CNN models with and without using their corresponding pretrained weights (PTWs) in the training for classifying $N$=100--299 ions (step: 1 ion) and for classifying $T$=5--15~mK (step: 1~mK) }
\end{figure}

In the above examples, the classification of $N$ and $T$ over certain ranges was demonstrated using the ResNet18 and AlexNet models, respectively, which were found to give the best performances in comparison with the other CNN models we evaluated. Figure \ref{fig:accuracies} shows the validation accuracies of all the selected models for classifying $N$ in the range 100--299 ions in steps of 1 ion and $T$ within 5--15~mK in steps of 1~mK. All models for classifying $N$ (or $T$) were trained on the same set of images with the same hyperparameters. In this figure, the validation accuracies of models without loading their respective PTWs in the training are also depicted. The classification performances generally decreased for both $N$ and $T$ when no PTWs were used, which confirms the merit of transfer learning. However, the extent to which the accuracies deteriorate vary by model. For example, the use of PTWs makes little difference in the performance of the VGG16 model, suggesting that it can still learn useful patterns from the training data even when starting from random weights. In contrast, the MnasNet1\_0 model showed a large drop in accuracy without PTWs. This is likely because MnasNet1\_0 is a smaller and more compact model designed to run efficiently on mobile or low-power devices. As a result, it has less capacity to learn complex patterns from scratch and therefore benefits more from being initialized with knowledge learned from a large dataset. For such models, starting with PTWs is especially important to achieve good performance. 

Figure \ref{fig:accuracies} also shows the importance of model selection to optimize the classification performance. We see that not only do different models perform quite differently for a specific task ($N$ or $T$), but the same model can also lead to very different results for different tasks. This also suggests that the VGG16 model is suitable for both tasks, achieving satisfactory performance. 

Note that in our examples, to achieve similarly high validation accuracy (93\% and 92\% for $N$ and $T$, respectively), almost six times more images (200 vs. 36) are required for each class for the classification of $T$ with 11 classes than for that of $N$ with 200 classes. Indeed, classifying $T$ appeared to be much more challenging than classifying $N$. For example, when we increased the range of $T$ to be classified from 5--15 mK (step:~1~mK) to 5--25 mK (step:~1~mK), the validation accuracy after convergence dropped to around 79\% with the AlexNet model under the same training conditions. Although $T$=5-15~mK is a typical temperature range of Coulomb crystals in practical situations \cite{mangeng2023}, we found that a wider range of 5--25~mK could be classified to a high accuracy of 98\% with ResNet18 if we could allow the step to be 2~mK, meaning that the model can classify images to temperature differences within 2~mK. For a comparison of classifying $T$=5--25~mK (step:~2 mK) with different CNN models, see Table \ref{tab:tab2}.

\subsection{Application to the analysis of experimental images}

\subsubsection{Transferring from simulations to experiments}
Once a CNN model is trained on simulated images, the weights of the model with the highest performance saved during training can be used in evaluation mode to classify other images of interest, whether they are from simulations or experiments. Although applying the model to simulated images can give a high classification accuracy, as suggested in Sec. \ref{subsec:examples}, its performance on experimental images may be significantly affected by the quality of the images. Experimental images can differ from simulated ones in terms of noise levels, distortions, contrast, etc., and they are susceptible to changes in experimental conditions. These differences pose challenges to the model's transferability from simulations to experiments. 

In spite of these challenges, we found that a few measures can be taken to alleviate the reduction of the model's performance when applied to experiments. First, the parameters used for generating the simulated training images should match or cover the experimental ones. The code used for this paper can easily be adapted to generate simulated images under different experimental parameters on which a new model can be trained. Second, when the experiment does not need to discern the variations of temperature or number of ions down to a precision of 1, the model can be trained on images with a higher variation. The classification accuracy should in general increase as the targeted precision is reduced, as can be inferred from Sec. \ref{subsec: model_comparison}. Last but not least, data augmentation \cite{krizhevsky2012imagenet, simonyan2014very, he2016deep, perez2017effectiveness, shorten2019survey} as an effective method developed in machine learning and computer vision can be employed before the images are fed into training. It can be used to increase the size and diversity of training datasets by applying various transformations to existing images, which can include rotations, scaling, noise injection, etc. This approach aims to improve the robustness and generalization ability of the machine-learned models, thus enhancing their performance on new, unseen images.

\begin{figure}[t]
    \includegraphics[width=\linewidth]{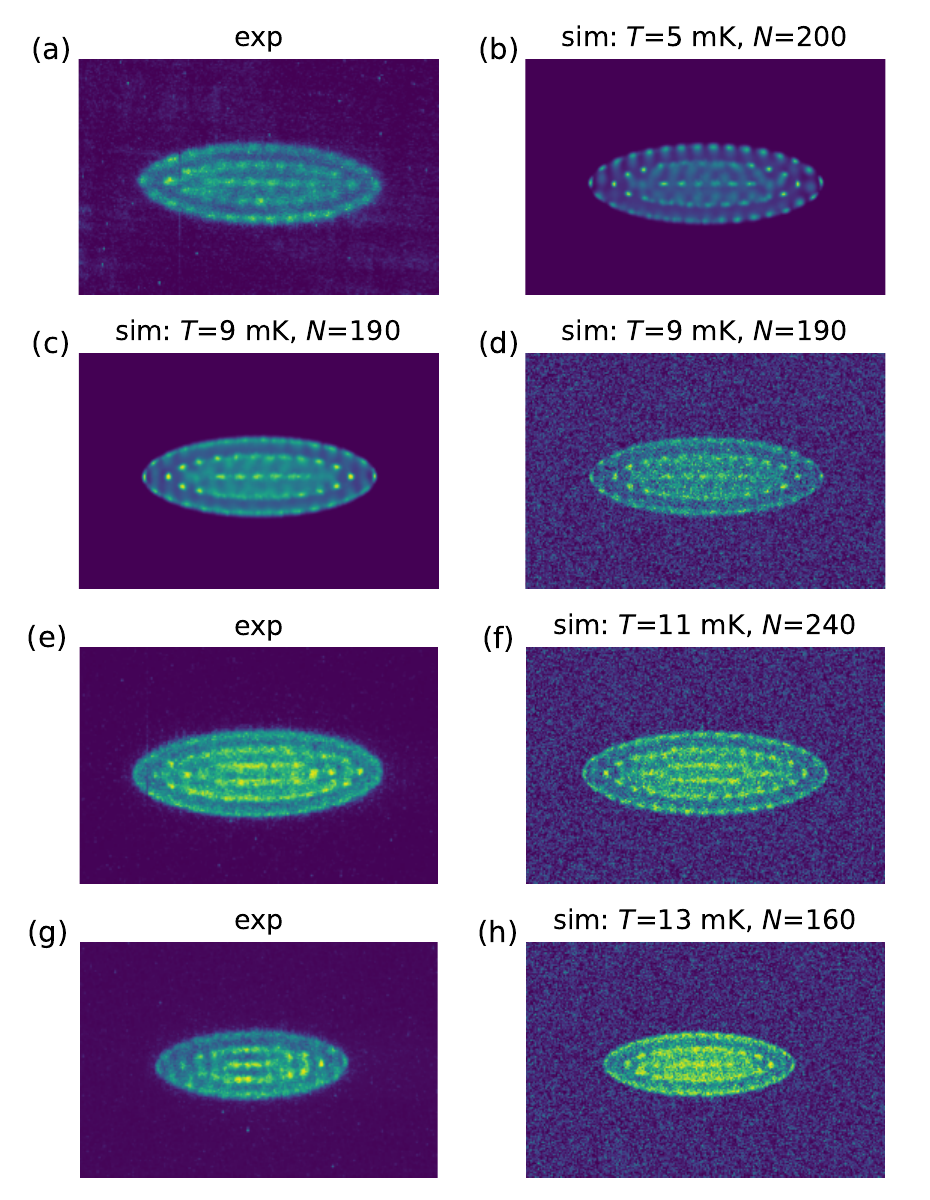}
    \caption{\label{fig:fig4} Comparison between experimental and simulated false-color images with $T$ and $N$ values predicted by CNN models using the experiment as inputs. 
    (a) First example experimental image. (b) Simulation of (a) with 200 ions at 5 mK. (c) Simulation of (a) with 190 ions at 9 mK. (d) The same simulated image as (c) but with added Gaussian noise. (e) Second example experimental image. (f) Simulation of (e) with 240 ions at 11 mK. (g) Third example experimental image. (h) Simulation of (g) with 160 ions at 13 mK.}
\end{figure}

Taking the experimental image shown in Fig. \ref{fig:fig4}(a) as an example, the AlexNet model trained on simulated images to classify $T$ in the range 5--25 mK in steps of 2 mK predicted a temperature of 5 mK. Meanwhile, the ResNet18 model trained to classify $N$ in the range 100--299 ions in larger steps of 10~ions predicted 200~ions. The corresponding simulated image with $T=5$~mK and $N=200$~ions is shown in Fig. \ref{fig:fig4}(b). Noticing that, in contrast to the simulated images, the experimental images contains noise, we performed data augmentation by adding Gaussian noise with mean $\mu=0$ and variance $\sigma^2=0.1$ grayscale intensity values (which was chosen after testing with a series of variances: 0.01, 0.05, 0.1, 0.15, 0.2, 0.25) to all the simulated images and then retrained the two models. This common data-augmentation strategy helps CNNs generalize better by narrowing the gap between clean simulated inputs and noisy experimental data. At the cost of a slightly reduced validation accuracy of 86\% with AlexNet, the thus-obtained model now predicted a temperature of 9 mK for the experimental image. Meanwhile, the retrained ResNet18 model with a validation accuracy of 87\% predicted 190 ions for the same image. A direct visual comparison of the corresponding noise-free and noise-augmented simulated images with $T=$ 9~mK and $N=$190 ions shown in Fig. \ref{fig:fig4}(c) and \ref{fig:fig4}(d), respectively, with the experiment in Fig. \ref{fig:fig4}(a), suggests that the new predictions provide a more realistic estimate of these attributes compared to the initial guess obtained without adding noise. Two additional examples are given in Fig. \ref{fig:fig4}(e-h), which compare the experimental images with the noise-added simulated images in which $T$ and $N$ are predicted by the retrained models. By visual comparison, we believe that the predicted values provide a fair representation of the experiments.

\subsubsection{Models trained directly on experimental images}

While we were unable to directly verify the output of the CNN models for large crystals with independent ion counting and thermometry methods (other than by visual comparison), we evaluated the generalizability and reliability of the present CNN approach using experimental images of Coulomb crystals containing small numbers of ions. Experimentally, we prepared a one-dimensional (1D) crystal with a defined number of ions (up to 12), and then transformed it to a 3D crystal by adjusting our trap voltages. To ensure that no ions were lost during the transformation, we repeatedly reverted the system to the 1D-crystal configuration before restoring it to the 3D case. Images of both cases with known labels of $N$ were saved and then used to train a CNN model to infer $N$.

\begin{figure}[t]
    \includegraphics[width=\linewidth]{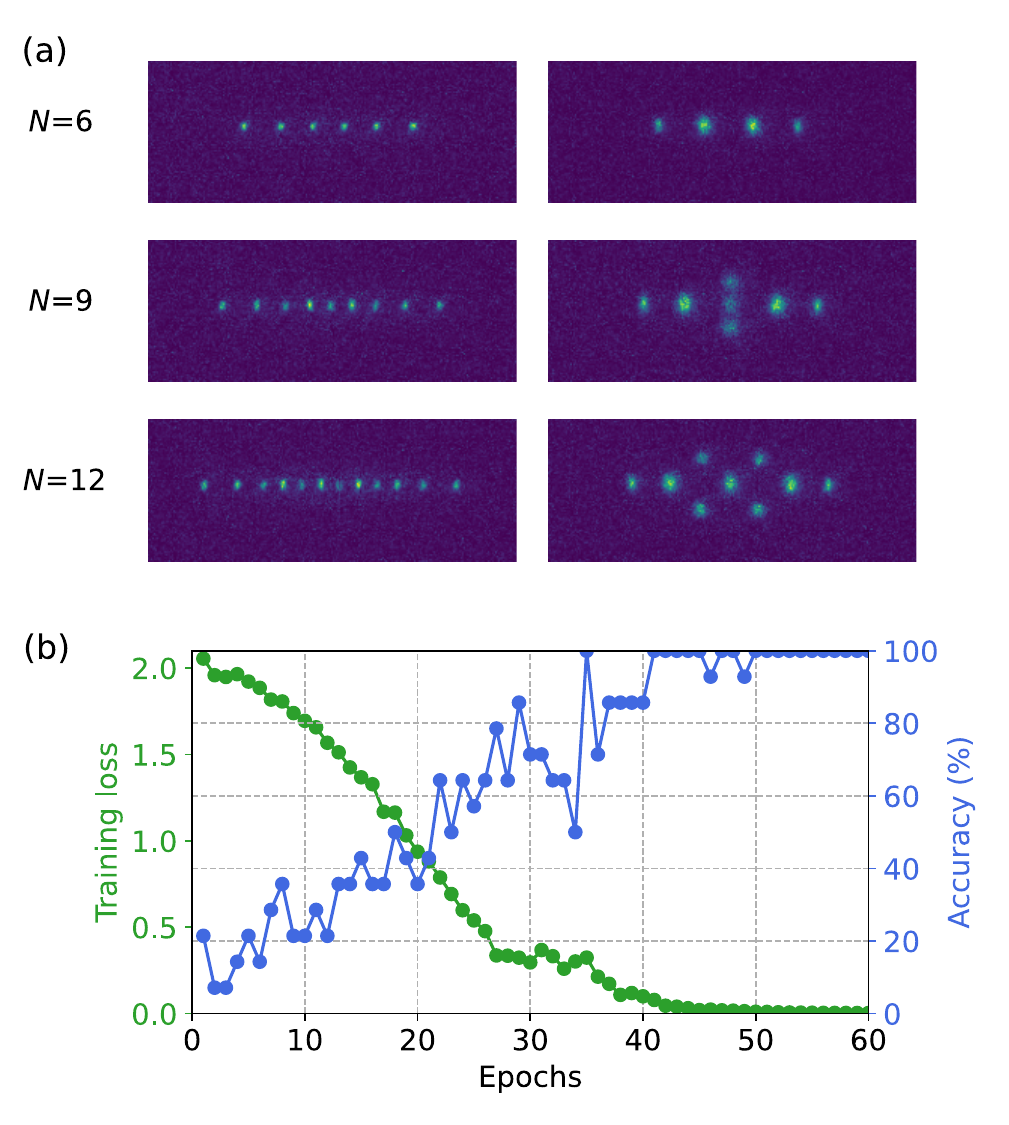}
    \caption{\label{fig:1d_3d_crystals}
    (a) Example experimental images (false color) of 1D (left) and 3D (right) Coulomb crystals with different $N$. (b) Training loss (in green) and validation accuracy (in blue) during training directly on experimental images.}
\end{figure}

We collected a total of 70 experimental images across seven classes of $N$ (4, 6, 7, 9, 10, 11, and 12), with each class containing five images of 1D crystals and five images of 3D crystals. Example images are shown in Fig.~\ref{fig:1d_3d_crystals}(a). After training a ResNet18 model with the same configurations as described in Sec. \ref{sec:method} on these images, the validation accuracy stabilized at 100\% after around 40 epochs. Figure~\ref{fig:1d_3d_crystals}(b) shows the evolution of the training loss and validation accuracy.

Given the small dataset size (70 images across seven classes), where random guessing would yield an accuracy of only $1/7\approx14.3$\%, the model's perfect accuracy demonstrates a strong learning capability even with limited real-world data, highlighting the extensibility and reliability of the approach.

\subsubsection{Potential for \emph{in-situ} and real-time inference }
In addition to applying the model \emph{a posteriori} to prerecorded experimental images, it can be directly integrated into an experiment's data-acquisition system, allowing an \emph{in-situ} determination of the relevant Coulomb-crystal parameters in real time. According to our evaluation results on a regular laboratory computer (CPU: Intel Core i7-3930K, RAM: 16 GB, no GPUs), the time taken for a model to classify an image with 512$\times$512 pixels is a few tens of milliseconds, sufficiently fast to capture changes in crystal characteristics in real time in typical experiments with trapped ions.

\section{\label{sec:conclusion} Conclusion and outlook}

We demonstrated a fast, accurate, and general methodology for determining the number and secular temperature of ions in Coulomb crystals using CNNs. This methodology involves the generation of a substantial dataset of images depicting crystals with varying parameters, specifically, the number of ions and their temperatures, through molecular-dynamics simulations. These images serve as a training set for CNN models tasked with classifying the aforementioned parameters. Upon evaluating the classification performance of several widely used pretrained CNN models, we identified ResNet18 and AlexNet as the most effective for classifying $N$ and $T$, respectively. Specifically, for crystals with an ion count ranging from 100 to 299 and temperatures between 5 and 15 mK, the best model achieves an accuracy of 93\% for the ion count and an accuracy of 92\% for the temperature when discerning variations as minute as one ion and 1 mK, respectively. 

With a slightly reduced accuracy, the CNN models trained on simulated images can be used for the classification of experimental images. We suggested several ways to improve the performance of experimental implementations, including reducing the classification precision and prepossessing the simulated input images (e.g., adding noise) before the training. Additionally, we demonstrated that CNN models trained directly on a small set of labeled experimental images could achieve high classification accuracy, highlighting the reliability and potential of the present approach. With a direct integration of the model into the experiments, real-time and \emph{in-situ} determination of crystal parameters is now feasible, thereby significantly enhancing the monitoring and control of trapped ions in such experiments.

We note that due to the limited number of experimental images with clearly defined numbers of ions and temperatures available for model evaluation, the actual performance of the models requires careful verification in real-life experimental settings. Clearly, the quality of the experimental image to be analyzed plays a critical role. Additionally, the manual classification of experimental images by visual comparison with simulated images, which are used to validate the model predictions, may be subject to inherent inaccuracies. Ideally, the models' predictions would be more reliable if they were trained on a large dataset of experimental images labeled using independent ion counting and thermometry methods, as mentioned in Sec. \ref{sec:intro}. When an insufficient number of experimental images is available, incorporating simulated images into a ``hybrid" experimental and simulated training dataset could be beneficial. However, if the models are trained partially or entirely on simulated images, potential systematic errors should be carefully calibrated before applying them to experiments. 

The present method is computationally efficient and enables fast data preparation, model training, and evaluation, while also being easily adaptable to different experimental parameters as needed. Furthermore, it can be readily adapted to other ion-trapping experiments, thereby enhancing the ion-counting and thermometry toolkit for trapped ions in the intermediate-temperature range. This compares with other parallel work on training a general model designed to accommodate a broader range of experimental conditions \cite{allsopp2025conv}. Potentially, it can also be extended to Coulomb crystals with mixed ions, facilitating studies with sympathetically cooled ions (see Refs. \cite{willitsch2012coulomb, hudson2016sympathetic} and references therein) and the dynamics of ion-neutral collisions \cite{willitsch2012coulomb,Tomza2019Cold, willitsch2015ion, eberle2015ion, dorfler2019long, deiss2024cold}.

\begin{acknowledgments}
We thank Prof. Timothy P. Softley (Birmingham) for helpful discussions. We thank Richard Karl and Christian Mangeng for their help with data collection. We acknowledge financial support from the Swiss National Science Foundation (Grants No. 200020\_175533, No. 200021\_204123, and No. TMAG-2\_209193), as well as the University of Basel. Y. Y. acknowledges support from the Research Fund of the University of Basel for Excellent Junior Researchers (Grant No. 3CH1051).

\end{acknowledgments}

\section*{Data availability}
The data that support the findings of this article are openly available \cite{yin2025code}. 

\bibliography{main}

\begin{thebibliography}{63}%
\makeatletter
\providecommand \@ifxundefined [1]{%
 \@ifx{#1\undefined}
}%
\providecommand \@ifnum [1]{%
 \ifnum #1\expandafter \@firstoftwo
 \else \expandafter \@secondoftwo
 \fi
}%
\providecommand \@ifx [1]{%
 \ifx #1\expandafter \@firstoftwo
 \else \expandafter \@secondoftwo
 \fi
}%
\providecommand \natexlab [1]{#1}%
\providecommand \enquote  [1]{``#1''}%
\providecommand \bibnamefont  [1]{#1}%
\providecommand \bibfnamefont [1]{#1}%
\providecommand \citenamefont [1]{#1}%
\providecommand \href@noop [0]{\@secondoftwo}%
\providecommand \href [0]{\begingroup \@sanitize@url \@href}%
\providecommand \@href[1]{\@@startlink{#1}\@@href}%
\providecommand \@@href[1]{\endgroup#1\@@endlink}%
\providecommand \@sanitize@url [0]{\catcode `\\12\catcode `\$12\catcode `\&12\catcode `\#12\catcode `\^12\catcode `\_12\catcode `\%12\relax}%
\providecommand \@@startlink[1]{}%
\providecommand \@@endlink[0]{}%
\providecommand \url  [0]{\begingroup\@sanitize@url \@url }%
\providecommand \@url [1]{\endgroup\@href {#1}{\urlprefix }}%
\providecommand \urlprefix  [0]{URL }%
\providecommand \Eprint [0]{\href }%
\providecommand \doibase [0]{https://doi.org/}%
\providecommand \selectlanguage [0]{\@gobble}%
\providecommand \bibinfo  [0]{\@secondoftwo}%
\providecommand \bibfield  [0]{\@secondoftwo}%
\providecommand \translation [1]{[#1]}%
\providecommand \BibitemOpen [0]{}%
\providecommand \bibitemStop [0]{}%
\providecommand \bibitemNoStop [0]{.\EOS\space}%
\providecommand \EOS [0]{\spacefactor3000\relax}%
\providecommand \BibitemShut  [1]{\csname bibitem#1\endcsname}%
\let\auto@bib@innerbib\@empty
\bibitem [{\citenamefont {Bruzewicz}\ \emph {et~al.}(2019)\citenamefont {Bruzewicz}, \citenamefont {Chiaverini}, \citenamefont {McConnell},\ and\ \citenamefont {Sage}}]{bruzewicz2019trapped}%
  \BibitemOpen
  \bibfield  {author} {\bibinfo {author} {\bibfnamefont {C.~D.}\ \bibnamefont {Bruzewicz}}, \bibinfo {author} {\bibfnamefont {J.}~\bibnamefont {Chiaverini}}, \bibinfo {author} {\bibfnamefont {R.}~\bibnamefont {McConnell}},\ and\ \bibinfo {author} {\bibfnamefont {J.~M.}\ \bibnamefont {Sage}},\ }\bibfield  {title} {\bibinfo {title} {Trapped-ion quantum computing: Progress and challenges},\ }\href {https://doi.org/10.1063/1.5088164} {\bibfield  {journal} {\bibinfo  {journal} {Appl. Phys. Rev.}\ }\textbf {\bibinfo {volume} {6}},\ \bibinfo {pages} {021314} (\bibinfo {year} {2019})}\BibitemShut {NoStop}%
\bibitem [{\citenamefont {Monroe}\ and\ \citenamefont {Kim}(2013)}]{monroe2013scaling}%
  \BibitemOpen
  \bibfield  {author} {\bibinfo {author} {\bibfnamefont {C.}~\bibnamefont {Monroe}}\ and\ \bibinfo {author} {\bibfnamefont {J.}~\bibnamefont {Kim}},\ }\bibfield  {title} {\bibinfo {title} {Scaling the ion trap quantum processor},\ }\href {https://doi.org/10.1126/science.1231298} {\bibfield  {journal} {\bibinfo  {journal} {Science}\ }\textbf {\bibinfo {volume} {339}},\ \bibinfo {pages} {1164} (\bibinfo {year} {2013})}\BibitemShut {NoStop}%
\bibitem [{\citenamefont {Häffner}\ \emph {et~al.}(2008)\citenamefont {Häffner}, \citenamefont {Roos},\ and\ \citenamefont {Blatt}}]{haffner2008quantum}%
  \BibitemOpen
  \bibfield  {author} {\bibinfo {author} {\bibfnamefont {H.}~\bibnamefont {Häffner}}, \bibinfo {author} {\bibfnamefont {C.}~\bibnamefont {Roos}},\ and\ \bibinfo {author} {\bibfnamefont {R.}~\bibnamefont {Blatt}},\ }\bibfield  {title} {\bibinfo {title} {Quantum computing with trapped ions},\ }\href {https://doi.org/https://doi.org/10.1016/j.physrep.2008.09.003} {\bibfield  {journal} {\bibinfo  {journal} {Phys. Rep.}\ }\textbf {\bibinfo {volume} {469}},\ \bibinfo {pages} {155} (\bibinfo {year} {2008})}\BibitemShut {NoStop}%
\bibitem [{\citenamefont {Wineland}(2009)}]{wineland2009quantum}%
  \BibitemOpen
  \bibfield  {author} {\bibinfo {author} {\bibfnamefont {D.~J.}\ \bibnamefont {Wineland}},\ }\bibfield  {title} {\bibinfo {title} {Quantum information processing and quantum control with trapped atomic ions},\ }\href {https://doi.org/10.1088/0031-8949/2009/T137/014007} {\bibfield  {journal} {\bibinfo  {journal} {Phys. Scr.}\ }\textbf {\bibinfo {volume} {2009}},\ \bibinfo {pages} {014007} (\bibinfo {year} {2009})}\BibitemShut {NoStop}%
\bibitem [{\citenamefont {Ludlow}\ \emph {et~al.}(2015)\citenamefont {Ludlow}, \citenamefont {Boyd}, \citenamefont {Ye}, \citenamefont {Peik},\ and\ \citenamefont {Schmidt}}]{ludlow2015optical}%
  \BibitemOpen
  \bibfield  {author} {\bibinfo {author} {\bibfnamefont {A.~D.}\ \bibnamefont {Ludlow}}, \bibinfo {author} {\bibfnamefont {M.~M.}\ \bibnamefont {Boyd}}, \bibinfo {author} {\bibfnamefont {J.}~\bibnamefont {Ye}}, \bibinfo {author} {\bibfnamefont {E.}~\bibnamefont {Peik}},\ and\ \bibinfo {author} {\bibfnamefont {P.~O.}\ \bibnamefont {Schmidt}},\ }\bibfield  {title} {\bibinfo {title} {Optical atomic clocks},\ }\href {https://doi.org/10.1103/RevModPhys.87.637} {\bibfield  {journal} {\bibinfo  {journal} {Rev. Mod. Phys.}\ }\textbf {\bibinfo {volume} {87}},\ \bibinfo {pages} {637} (\bibinfo {year} {2015})}\BibitemShut {NoStop}%
\bibitem [{\citenamefont {Safronova}\ \emph {et~al.}(2018)\citenamefont {Safronova}, \citenamefont {Budker}, \citenamefont {DeMille}, \citenamefont {Kimball}, \citenamefont {Derevianko},\ and\ \citenamefont {Clark}}]{safronova2018search}%
  \BibitemOpen
  \bibfield  {author} {\bibinfo {author} {\bibfnamefont {M.~S.}\ \bibnamefont {Safronova}}, \bibinfo {author} {\bibfnamefont {D.}~\bibnamefont {Budker}}, \bibinfo {author} {\bibfnamefont {D.}~\bibnamefont {DeMille}}, \bibinfo {author} {\bibfnamefont {D.~F.~J.}\ \bibnamefont {Kimball}}, \bibinfo {author} {\bibfnamefont {A.}~\bibnamefont {Derevianko}},\ and\ \bibinfo {author} {\bibfnamefont {C.~W.}\ \bibnamefont {Clark}},\ }\bibfield  {title} {\bibinfo {title} {Search for new physics with atoms and molecules},\ }\href {https://doi.org/10.1103/RevModPhys.90.025008} {\bibfield  {journal} {\bibinfo  {journal} {Rev. Mod. Phys.}\ }\textbf {\bibinfo {volume} {90}},\ \bibinfo {pages} {025008} (\bibinfo {year} {2018})}\BibitemShut {NoStop}%
\bibitem [{\citenamefont {Willitsch}(2012)}]{willitsch2012coulomb}%
  \BibitemOpen
  \bibfield  {author} {\bibinfo {author} {\bibfnamefont {S.}~\bibnamefont {Willitsch}},\ }\bibfield  {title} {\bibinfo {title} {Coulomb-crystallised molecular ions in traps: methods, applications, prospects},\ }\href {https://doi.org/10.1080/0144235X.2012.667221} {\bibfield  {journal} {\bibinfo  {journal} {Int. Rev. Phys. Chem.}\ }\textbf {\bibinfo {volume} {31}},\ \bibinfo {pages} {175} (\bibinfo {year} {2012})}\BibitemShut {NoStop}%
\bibitem [{\citenamefont {Major}\ \emph {et~al.}(2005)\citenamefont {Major}, \citenamefont {Gheorghe},\ and\ \citenamefont {Werth}}]{major2005charged}%
  \BibitemOpen
  \bibfield  {author} {\bibinfo {author} {\bibfnamefont {F.~G.}\ \bibnamefont {Major}}, \bibinfo {author} {\bibfnamefont {V.~N.}\ \bibnamefont {Gheorghe}},\ and\ \bibinfo {author} {\bibfnamefont {G.}~\bibnamefont {Werth}},\ }\href@noop {} {\emph {\bibinfo {title} {Charged particle traps: physics and techniques of charged particle field confinement}}},\ Vol.~\bibinfo {volume} {37}\ (\bibinfo  {publisher} {Springer Science \& Business Media},\ \bibinfo {year} {2005})\BibitemShut {NoStop}%
\bibitem [{\citenamefont {Willitsch}\ \emph {et~al.}(2008)\citenamefont {Willitsch}, \citenamefont {Bell}, \citenamefont {Gingell},\ and\ \citenamefont {Softley}}]{willitsch08b}%
  \BibitemOpen
  \bibfield  {author} {\bibinfo {author} {\bibfnamefont {S.}~\bibnamefont {Willitsch}}, \bibinfo {author} {\bibfnamefont {M.~T.}\ \bibnamefont {Bell}}, \bibinfo {author} {\bibfnamefont {A.~D.}\ \bibnamefont {Gingell}},\ and\ \bibinfo {author} {\bibfnamefont {T.~P.}\ \bibnamefont {Softley}},\ }\bibfield  {title} {\bibinfo {title} {Chemical applications of laser- and sympathetically-cooled ions in ion traps},\ }\href {https://doi.org/https://doi.org/10.1039/B813408C} {\bibfield  {journal} {\bibinfo  {journal} {Phys. Chem. Chem. Phys.}\ }\textbf {\bibinfo {volume} {10}},\ \bibinfo {pages} {7200} (\bibinfo {year} {2008})}\BibitemShut {NoStop}%
\bibitem [{\citenamefont {Schmid}\ \emph {et~al.}(2022)\citenamefont {Schmid}, \citenamefont {Weitenberg}, \citenamefont {Moreno}, \citenamefont {H\"ansch}, \citenamefont {Udem},\ and\ \citenamefont {Ozawa}}]{Schmid2022number}%
  \BibitemOpen
  \bibfield  {author} {\bibinfo {author} {\bibfnamefont {F.}~\bibnamefont {Schmid}}, \bibinfo {author} {\bibfnamefont {J.}~\bibnamefont {Weitenberg}}, \bibinfo {author} {\bibfnamefont {J.}~\bibnamefont {Moreno}}, \bibinfo {author} {\bibfnamefont {T.~W.}\ \bibnamefont {H\"ansch}}, \bibinfo {author} {\bibfnamefont {T.}~\bibnamefont {Udem}},\ and\ \bibinfo {author} {\bibfnamefont {A.}~\bibnamefont {Ozawa}},\ }\bibfield  {title} {\bibinfo {title} {Number-resolved detection of dark ions in coulomb crystals},\ }\href {https://doi.org/10.1103/PhysRevA.106.L041101} {\bibfield  {journal} {\bibinfo  {journal} {Phys. Rev. A}\ }\textbf {\bibinfo {volume} {106}},\ \bibinfo {pages} {L041101} (\bibinfo {year} {2022})}\BibitemShut {NoStop}%
\bibitem [{\citenamefont {Schneider}\ \emph {et~al.}(2014)\citenamefont {Schneider}, \citenamefont {Schowalter}, \citenamefont {Chen}, \citenamefont {Sullivan},\ and\ \citenamefont {Hudson}}]{schneider2014laser}%
  \BibitemOpen
  \bibfield  {author} {\bibinfo {author} {\bibfnamefont {C.}~\bibnamefont {Schneider}}, \bibinfo {author} {\bibfnamefont {S.~J.}\ \bibnamefont {Schowalter}}, \bibinfo {author} {\bibfnamefont {K.}~\bibnamefont {Chen}}, \bibinfo {author} {\bibfnamefont {S.~T.}\ \bibnamefont {Sullivan}},\ and\ \bibinfo {author} {\bibfnamefont {E.~R.}\ \bibnamefont {Hudson}},\ }\bibfield  {title} {\bibinfo {title} {Laser-cooling-assisted mass spectrometry},\ }\href {https://doi.org/10.1103/PhysRevApplied.2.034013} {\bibfield  {journal} {\bibinfo  {journal} {Phys. Rev. Appl.}\ }\textbf {\bibinfo {volume} {2}},\ \bibinfo {pages} {034013} (\bibinfo {year} {2014})}\BibitemShut {NoStop}%
\bibitem [{\citenamefont {Meyer}\ \emph {et~al.}(2015)\citenamefont {Meyer}, \citenamefont {Pollum}, \citenamefont {Petralia}, \citenamefont {Tauschinsky}, \citenamefont {Rennick}, \citenamefont {Softley},\ and\ \citenamefont {Heazlewood}}]{meyer2015ejection}%
  \BibitemOpen
  \bibfield  {author} {\bibinfo {author} {\bibfnamefont {K.~A.~E.}\ \bibnamefont {Meyer}}, \bibinfo {author} {\bibfnamefont {L.~L.}\ \bibnamefont {Pollum}}, \bibinfo {author} {\bibfnamefont {L.~S.}\ \bibnamefont {Petralia}}, \bibinfo {author} {\bibfnamefont {A.}~\bibnamefont {Tauschinsky}}, \bibinfo {author} {\bibfnamefont {C.~J.}\ \bibnamefont {Rennick}}, \bibinfo {author} {\bibfnamefont {T.~P.}\ \bibnamefont {Softley}},\ and\ \bibinfo {author} {\bibfnamefont {B.~R.}\ \bibnamefont {Heazlewood}},\ }\bibfield  {title} {\bibinfo {title} {Ejection of coulomb crystals from a linear paul ion trap for ion–molecule reaction studies},\ }\href {https://doi.org/10.1021/acs.jpca.5b07919} {\bibfield  {journal} {\bibinfo  {journal} {J. Phys. Chem. A}\ }\textbf {\bibinfo {volume} {119}},\ \bibinfo {pages} {12449} (\bibinfo {year} {2015})}\BibitemShut {NoStop}%
\bibitem [{\citenamefont {Schmid}\ \emph {et~al.}(2017)\citenamefont {Schmid}, \citenamefont {Greenberg}, \citenamefont {Miller}, \citenamefont {Loeffler},\ and\ \citenamefont {Lewandowski}}]{schmid2017ion}%
  \BibitemOpen
  \bibfield  {author} {\bibinfo {author} {\bibfnamefont {P.~C.}\ \bibnamefont {Schmid}}, \bibinfo {author} {\bibfnamefont {J.}~\bibnamefont {Greenberg}}, \bibinfo {author} {\bibfnamefont {M.~I.}\ \bibnamefont {Miller}}, \bibinfo {author} {\bibfnamefont {K.}~\bibnamefont {Loeffler}},\ and\ \bibinfo {author} {\bibfnamefont {H.~J.}\ \bibnamefont {Lewandowski}},\ }\bibfield  {title} {\bibinfo {title} {An ion trap time-of-flight mass spectrometer with high mass resolution for cold trapped ion experiments},\ }\href {https://doi.org/10.1063/1.4996911} {\bibfield  {journal} {\bibinfo  {journal} {Rev. Sci. Instrum.}\ }\textbf {\bibinfo {volume} {88}},\ \bibinfo {pages} {123107} (\bibinfo {year} {2017})}\BibitemShut {NoStop}%
\bibitem [{\citenamefont {R\"osch}\ \emph {et~al.}(2016)\citenamefont {R\"osch}, \citenamefont {Gao}, \citenamefont {Kilaj},\ and\ \citenamefont {Willitsch}}]{roesch16a}%
  \BibitemOpen
  \bibfield  {author} {\bibinfo {author} {\bibfnamefont {D.}~\bibnamefont {R\"osch}}, \bibinfo {author} {\bibfnamefont {H.}~\bibnamefont {Gao}}, \bibinfo {author} {\bibfnamefont {A.}~\bibnamefont {Kilaj}},\ and\ \bibinfo {author} {\bibfnamefont {S.}~\bibnamefont {Willitsch}},\ }\bibfield  {title} {\bibinfo {title} {Design and characterization of a linear quadrupole ion trap for high-resolution coulomb-crystal time-of-flight mass spectrometry},\ }\href@noop {} {\bibfield  {journal} {\bibinfo  {journal} {EPJ Tech. Instrum.}\ }\textbf {\bibinfo {volume} {3}},\ \bibinfo {pages} {5} (\bibinfo {year} {2016})}\BibitemShut {NoStop}%
\bibitem [{\citenamefont {Herrmann}\ \emph {et~al.}(2009)\citenamefont {Herrmann}, \citenamefont {Batteiger}, \citenamefont {Kn\"unz}, \citenamefont {Saathoff}, \citenamefont {Udem},\ and\ \citenamefont {H\"ansch}}]{Herrmann2009Frequency}%
  \BibitemOpen
  \bibfield  {author} {\bibinfo {author} {\bibfnamefont {M.}~\bibnamefont {Herrmann}}, \bibinfo {author} {\bibfnamefont {V.}~\bibnamefont {Batteiger}}, \bibinfo {author} {\bibfnamefont {S.}~\bibnamefont {Kn\"unz}}, \bibinfo {author} {\bibfnamefont {G.}~\bibnamefont {Saathoff}}, \bibinfo {author} {\bibfnamefont {T.}~\bibnamefont {Udem}},\ and\ \bibinfo {author} {\bibfnamefont {T.~W.}\ \bibnamefont {H\"ansch}},\ }\bibfield  {title} {\bibinfo {title} {Frequency metrology on single trapped ions in the weak binding limit: The $3{s}_{1/2}\ensuremath{-}3{p}_{3/2}$ transition in $^{24}\mathrm{Mg}^{+}$},\ }\href {https://doi.org/10.1103/PhysRevLett.102.013006} {\bibfield  {journal} {\bibinfo  {journal} {Phys. Rev. Lett.}\ }\textbf {\bibinfo {volume} {102}},\ \bibinfo {pages} {013006} (\bibinfo {year} {2009})}\BibitemShut {NoStop}%
\bibitem [{\citenamefont {Roßnagel}\ \emph {et~al.}(2015)\citenamefont {Roßnagel}, \citenamefont {Tolazzi}, \citenamefont {Schmidt-Kaler},\ and\ \citenamefont {Singer}}]{Roßnagel2015fast}%
  \BibitemOpen
  \bibfield  {author} {\bibinfo {author} {\bibfnamefont {J.}~\bibnamefont {Roßnagel}}, \bibinfo {author} {\bibfnamefont {K.~N.}\ \bibnamefont {Tolazzi}}, \bibinfo {author} {\bibfnamefont {F.}~\bibnamefont {Schmidt-Kaler}},\ and\ \bibinfo {author} {\bibfnamefont {K.}~\bibnamefont {Singer}},\ }\bibfield  {title} {\bibinfo {title} {Fast thermometry for trapped ions using dark resonances},\ }\href {https://doi.org/10.1088/1367-2630/17/4/045004} {\bibfield  {journal} {\bibinfo  {journal} {New J. Phys.}\ }\textbf {\bibinfo {volume} {17}},\ \bibinfo {pages} {045004} (\bibinfo {year} {2015})}\BibitemShut {NoStop}%
\bibitem [{\citenamefont {Norton}\ \emph {et~al.}(2011)\citenamefont {Norton}, \citenamefont {Streed}, \citenamefont {Petrasiunas}, \citenamefont {Jechow},\ and\ \citenamefont {Kielpinski}}]{norton2011millikelvin}%
  \BibitemOpen
  \bibfield  {author} {\bibinfo {author} {\bibfnamefont {B.~G.}\ \bibnamefont {Norton}}, \bibinfo {author} {\bibfnamefont {E.~W.}\ \bibnamefont {Streed}}, \bibinfo {author} {\bibfnamefont {M.~J.}\ \bibnamefont {Petrasiunas}}, \bibinfo {author} {\bibfnamefont {A.}~\bibnamefont {Jechow}},\ and\ \bibinfo {author} {\bibfnamefont {D.}~\bibnamefont {Kielpinski}},\ }\bibfield  {title} {\bibinfo {title} {Millikelvin spatial thermometry of trapped ions},\ }\href {https://doi.org/10.1088/1367-2630/13/11/113022} {\bibfield  {journal} {\bibinfo  {journal} {New J. Phys.}\ }\textbf {\bibinfo {volume} {13}},\ \bibinfo {pages} {113022} (\bibinfo {year} {2011})}\BibitemShut {NoStop}%
\bibitem [{\citenamefont {Kn\"unz}\ \emph {et~al.}(2012)\citenamefont {Kn\"unz}, \citenamefont {Herrmann}, \citenamefont {Batteiger}, \citenamefont {Saathoff}, \citenamefont {H\"ansch},\ and\ \citenamefont {Udem}}]{knunz2012sub}%
  \BibitemOpen
  \bibfield  {author} {\bibinfo {author} {\bibfnamefont {S.}~\bibnamefont {Kn\"unz}}, \bibinfo {author} {\bibfnamefont {M.}~\bibnamefont {Herrmann}}, \bibinfo {author} {\bibfnamefont {V.}~\bibnamefont {Batteiger}}, \bibinfo {author} {\bibfnamefont {G.}~\bibnamefont {Saathoff}}, \bibinfo {author} {\bibfnamefont {T.~W.}\ \bibnamefont {H\"ansch}},\ and\ \bibinfo {author} {\bibfnamefont {T.}~\bibnamefont {Udem}},\ }\bibfield  {title} {\bibinfo {title} {Sub-millikelvin spatial thermometry of a single doppler-cooled ion in a paul trap},\ }\href {https://doi.org/10.1103/PhysRevA.85.023427} {\bibfield  {journal} {\bibinfo  {journal} {Phys. Rev. A}\ }\textbf {\bibinfo {volume} {85}},\ \bibinfo {pages} {023427} (\bibinfo {year} {2012})}\BibitemShut {NoStop}%
\bibitem [{\citenamefont {Rajagopal}\ \emph {et~al.}(2016)\citenamefont {Rajagopal}, \citenamefont {Marler}, \citenamefont {Kokish},\ and\ \citenamefont {Odom}}]{rajagopal2016trapped}%
  \BibitemOpen
  \bibfield  {author} {\bibinfo {author} {\bibfnamefont {V.}~\bibnamefont {Rajagopal}}, \bibinfo {author} {\bibfnamefont {J.~P.}\ \bibnamefont {Marler}}, \bibinfo {author} {\bibfnamefont {M.~G.}\ \bibnamefont {Kokish}},\ and\ \bibinfo {author} {\bibfnamefont {B.~C.}\ \bibnamefont {Odom}},\ }\bibfield  {title} {\bibinfo {title} {Trapped ion chain thermometry and mass spectrometry through imaging},\ }\href {https://doi.org/10.1255/ejms.1408} {\bibfield  {journal} {\bibinfo  {journal} {Eur. J. Mass Spectrom.}\ }\textbf {\bibinfo {volume} {22}},\ \bibinfo {pages} {1} (\bibinfo {year} {2016})}\BibitemShut {NoStop}%
\bibitem [{\citenamefont {Prestage}\ \emph {et~al.}(1991)\citenamefont {Prestage}, \citenamefont {Williams}, \citenamefont {Maleki}, \citenamefont {Djomehri},\ and\ \citenamefont {Harabetian}}]{Prestage1991}%
  \BibitemOpen
  \bibfield  {author} {\bibinfo {author} {\bibfnamefont {J.~D.}\ \bibnamefont {Prestage}}, \bibinfo {author} {\bibfnamefont {A.}~\bibnamefont {Williams}}, \bibinfo {author} {\bibfnamefont {L.}~\bibnamefont {Maleki}}, \bibinfo {author} {\bibfnamefont {M.~J.}\ \bibnamefont {Djomehri}},\ and\ \bibinfo {author} {\bibfnamefont {E.}~\bibnamefont {Harabetian}},\ }\bibfield  {title} {\bibinfo {title} {Dynamics of charged particles in a paul radio-frequency quadrupole trap},\ }\href {https://doi.org/10.1103/PhysRevLett.66.2964} {\bibfield  {journal} {\bibinfo  {journal} {Phys. Rev. Lett.}\ }\textbf {\bibinfo {volume} {66}},\ \bibinfo {pages} {2964} (\bibinfo {year} {1991})}\BibitemShut {NoStop}%
\bibitem [{\citenamefont {Ostendorf}\ \emph {et~al.}(2006)\citenamefont {Ostendorf}, \citenamefont {Zhang}, \citenamefont {Wilson}, \citenamefont {Offenberg}, \citenamefont {Roth},\ and\ \citenamefont {Schiller}}]{Ostendorf2006}%
  \BibitemOpen
  \bibfield  {author} {\bibinfo {author} {\bibfnamefont {A.}~\bibnamefont {Ostendorf}}, \bibinfo {author} {\bibfnamefont {C.~B.}\ \bibnamefont {Zhang}}, \bibinfo {author} {\bibfnamefont {M.~A.}\ \bibnamefont {Wilson}}, \bibinfo {author} {\bibfnamefont {D.}~\bibnamefont {Offenberg}}, \bibinfo {author} {\bibfnamefont {B.}~\bibnamefont {Roth}},\ and\ \bibinfo {author} {\bibfnamefont {S.}~\bibnamefont {Schiller}},\ }\bibfield  {title} {\bibinfo {title} {Sympathetic cooling of complex molecular ions to millikelvin temperatures},\ }\href {https://doi.org/10.1103/PhysRevLett.97.243005} {\bibfield  {journal} {\bibinfo  {journal} {Phys. Rev. Lett.}\ }\textbf {\bibinfo {volume} {97}},\ \bibinfo {pages} {243005} (\bibinfo {year} {2006})}\BibitemShut {NoStop}%
\bibitem [{\citenamefont {Zhang}\ \emph {et~al.}(2007)\citenamefont {Zhang}, \citenamefont {Offenberg}, \citenamefont {Roth}, \citenamefont {Wilson},\ and\ \citenamefont {Schiller}}]{Zhang2007}%
  \BibitemOpen
  \bibfield  {author} {\bibinfo {author} {\bibfnamefont {C.~B.}\ \bibnamefont {Zhang}}, \bibinfo {author} {\bibfnamefont {D.}~\bibnamefont {Offenberg}}, \bibinfo {author} {\bibfnamefont {B.}~\bibnamefont {Roth}}, \bibinfo {author} {\bibfnamefont {M.~A.}\ \bibnamefont {Wilson}},\ and\ \bibinfo {author} {\bibfnamefont {S.}~\bibnamefont {Schiller}},\ }\bibfield  {title} {\bibinfo {title} {Molecular-dynamics simulations of cold single-species and multispecies ion ensembles in a linear paul trap},\ }\href {https://doi.org/10.1103/PhysRevA.76.012719} {\bibfield  {journal} {\bibinfo  {journal} {Phys. Rev. A}\ }\textbf {\bibinfo {volume} {76}},\ \bibinfo {pages} {012719} (\bibinfo {year} {2007})}\BibitemShut {NoStop}%
\bibitem [{\citenamefont {Okada}\ \emph {et~al.}(2010)\citenamefont {Okada}, \citenamefont {Wada}, \citenamefont {Takayanagi}, \citenamefont {Ohtani},\ and\ \citenamefont {Schuessler}}]{Okada2010}%
  \BibitemOpen
  \bibfield  {author} {\bibinfo {author} {\bibfnamefont {K.}~\bibnamefont {Okada}}, \bibinfo {author} {\bibfnamefont {M.}~\bibnamefont {Wada}}, \bibinfo {author} {\bibfnamefont {T.}~\bibnamefont {Takayanagi}}, \bibinfo {author} {\bibfnamefont {S.}~\bibnamefont {Ohtani}},\ and\ \bibinfo {author} {\bibfnamefont {H.~A.}\ \bibnamefont {Schuessler}},\ }\bibfield  {title} {\bibinfo {title} {Characterization of ion coulomb crystals in a linear paul trap},\ }\href {https://doi.org/10.1103/PhysRevA.81.013420} {\bibfield  {journal} {\bibinfo  {journal} {Phys. Rev. A}\ }\textbf {\bibinfo {volume} {81}},\ \bibinfo {pages} {013420} (\bibinfo {year} {2010})}\BibitemShut {NoStop}%
\bibitem [{\citenamefont {Tong}\ \emph {et~al.}(2010)\citenamefont {Tong}, \citenamefont {Winney},\ and\ \citenamefont {Willitsch}}]{Tong2010}%
  \BibitemOpen
  \bibfield  {author} {\bibinfo {author} {\bibfnamefont {X.}~\bibnamefont {Tong}}, \bibinfo {author} {\bibfnamefont {A.~H.}\ \bibnamefont {Winney}},\ and\ \bibinfo {author} {\bibfnamefont {S.}~\bibnamefont {Willitsch}},\ }\bibfield  {title} {\bibinfo {title} {Sympathetic cooling of molecular ions in selected rotational and vibrational states produced by threshold photoionization},\ }\href {https://doi.org/10.1103/PhysRevLett.105.143001} {\bibfield  {journal} {\bibinfo  {journal} {Phys. Rev. Lett.}\ }\textbf {\bibinfo {volume} {105}},\ \bibinfo {pages} {143001} (\bibinfo {year} {2010})}\BibitemShut {NoStop}%
\bibitem [{\citenamefont {Chen}\ \emph {et~al.}(2020)\citenamefont {Chen}, \citenamefont {Wright}, \citenamefont {Pisenti}, \citenamefont {Murphy}, \citenamefont {Beck}, \citenamefont {Landsman}, \citenamefont {Amini},\ and\ \citenamefont {Nam}}]{Chen2020Efficient}%
  \BibitemOpen
  \bibfield  {author} {\bibinfo {author} {\bibfnamefont {J.-S.}\ \bibnamefont {Chen}}, \bibinfo {author} {\bibfnamefont {K.}~\bibnamefont {Wright}}, \bibinfo {author} {\bibfnamefont {N.~C.}\ \bibnamefont {Pisenti}}, \bibinfo {author} {\bibfnamefont {D.}~\bibnamefont {Murphy}}, \bibinfo {author} {\bibfnamefont {K.~M.}\ \bibnamefont {Beck}}, \bibinfo {author} {\bibfnamefont {K.}~\bibnamefont {Landsman}}, \bibinfo {author} {\bibfnamefont {J.~M.}\ \bibnamefont {Amini}},\ and\ \bibinfo {author} {\bibfnamefont {Y.}~\bibnamefont {Nam}},\ }\bibfield  {title} {\bibinfo {title} {Efficient-sideband-cooling protocol for long trapped-ion chains},\ }\href {https://doi.org/10.1103/PhysRevA.102.043110} {\bibfield  {journal} {\bibinfo  {journal} {Phys. Rev. A}\ }\textbf {\bibinfo {volume} {102}},\ \bibinfo {pages} {043110} (\bibinfo {year} {2020})}\BibitemShut {NoStop}%
\bibitem [{\citenamefont {Feng}\ \emph {et~al.}(2020)\citenamefont {Feng}, \citenamefont {Tan}, \citenamefont {De}, \citenamefont {Menon}, \citenamefont {Chu}, \citenamefont {Pagano},\ and\ \citenamefont {Monroe}}]{Feng2020Efficient}%
  \BibitemOpen
  \bibfield  {author} {\bibinfo {author} {\bibfnamefont {L.}~\bibnamefont {Feng}}, \bibinfo {author} {\bibfnamefont {W.~L.}\ \bibnamefont {Tan}}, \bibinfo {author} {\bibfnamefont {A.}~\bibnamefont {De}}, \bibinfo {author} {\bibfnamefont {A.}~\bibnamefont {Menon}}, \bibinfo {author} {\bibfnamefont {A.}~\bibnamefont {Chu}}, \bibinfo {author} {\bibfnamefont {G.}~\bibnamefont {Pagano}},\ and\ \bibinfo {author} {\bibfnamefont {C.}~\bibnamefont {Monroe}},\ }\bibfield  {title} {\bibinfo {title} {Efficient ground-state cooling of large trapped-ion chains with an electromagnetically-induced-transparency tripod scheme},\ }\href {https://doi.org/10.1103/PhysRevLett.125.053001} {\bibfield  {journal} {\bibinfo  {journal} {Phys. Rev. Lett.}\ }\textbf {\bibinfo {volume} {125}},\ \bibinfo {pages} {053001} (\bibinfo {year} {2020})}\BibitemShut {NoStop}%
\bibitem [{\citenamefont {Sawyer}\ \emph {et~al.}(2012)\citenamefont {Sawyer}, \citenamefont {Britton}, \citenamefont {Keith}, \citenamefont {Wang}, \citenamefont {Freericks}, \citenamefont {Uys}, \citenamefont {Biercuk},\ and\ \citenamefont {Bollinger}}]{Sawyer2012Spectroscopy}%
  \BibitemOpen
  \bibfield  {author} {\bibinfo {author} {\bibfnamefont {B.~C.}\ \bibnamefont {Sawyer}}, \bibinfo {author} {\bibfnamefont {J.~W.}\ \bibnamefont {Britton}}, \bibinfo {author} {\bibfnamefont {A.~C.}\ \bibnamefont {Keith}}, \bibinfo {author} {\bibfnamefont {C.-C.~J.}\ \bibnamefont {Wang}}, \bibinfo {author} {\bibfnamefont {J.~K.}\ \bibnamefont {Freericks}}, \bibinfo {author} {\bibfnamefont {H.}~\bibnamefont {Uys}}, \bibinfo {author} {\bibfnamefont {M.~J.}\ \bibnamefont {Biercuk}},\ and\ \bibinfo {author} {\bibfnamefont {J.~J.}\ \bibnamefont {Bollinger}},\ }\bibfield  {title} {\bibinfo {title} {Spectroscopy and thermometry of drumhead modes in a mesoscopic trapped-ion crystal using entanglement},\ }\href {https://doi.org/10.1103/PhysRevLett.108.213003} {\bibfield  {journal} {\bibinfo  {journal} {Phys. Rev. Lett.}\ }\textbf {\bibinfo {volume} {108}},\ \bibinfo {pages} {213003} (\bibinfo {year} {2012})}\BibitemShut {NoStop}%
\bibitem [{\citenamefont {D'Onofrio}\ \emph {et~al.}(2021)\citenamefont {D'Onofrio}, \citenamefont {Xie}, \citenamefont {Rasmusson}, \citenamefont {Wolanski}, \citenamefont {Cui},\ and\ \citenamefont {Richerme}}]{DOnofrio2021}%
  \BibitemOpen
  \bibfield  {author} {\bibinfo {author} {\bibfnamefont {M.}~\bibnamefont {D'Onofrio}}, \bibinfo {author} {\bibfnamefont {Y.}~\bibnamefont {Xie}}, \bibinfo {author} {\bibfnamefont {A.~J.}\ \bibnamefont {Rasmusson}}, \bibinfo {author} {\bibfnamefont {E.}~\bibnamefont {Wolanski}}, \bibinfo {author} {\bibfnamefont {J.}~\bibnamefont {Cui}},\ and\ \bibinfo {author} {\bibfnamefont {P.}~\bibnamefont {Richerme}},\ }\bibfield  {title} {\bibinfo {title} {Radial two-dimensional ion crystals in a linear paul trap},\ }\href {https://doi.org/10.1103/PhysRevLett.127.020503} {\bibfield  {journal} {\bibinfo  {journal} {Phys. Rev. Lett.}\ }\textbf {\bibinfo {volume} {127}},\ \bibinfo {pages} {020503} (\bibinfo {year} {2021})}\BibitemShut {NoStop}%
\bibitem [{\citenamefont {Vybornyi}\ \emph {et~al.}(2023)\citenamefont {Vybornyi}, \citenamefont {Dreissen}, \citenamefont {Kiesenhofer}, \citenamefont {Hainzer}, \citenamefont {Bock}, \citenamefont {Ollikainen}, \citenamefont {Vadlejch}, \citenamefont {Roos}, \citenamefont {Mehlst\"aubler},\ and\ \citenamefont {Hammerer}}]{Vybornyi2023Sideband}%
  \BibitemOpen
  \bibfield  {author} {\bibinfo {author} {\bibfnamefont {I.}~\bibnamefont {Vybornyi}}, \bibinfo {author} {\bibfnamefont {L.~S.}\ \bibnamefont {Dreissen}}, \bibinfo {author} {\bibfnamefont {D.}~\bibnamefont {Kiesenhofer}}, \bibinfo {author} {\bibfnamefont {H.}~\bibnamefont {Hainzer}}, \bibinfo {author} {\bibfnamefont {M.}~\bibnamefont {Bock}}, \bibinfo {author} {\bibfnamefont {T.}~\bibnamefont {Ollikainen}}, \bibinfo {author} {\bibfnamefont {D.}~\bibnamefont {Vadlejch}}, \bibinfo {author} {\bibfnamefont {C.~F.}\ \bibnamefont {Roos}}, \bibinfo {author} {\bibfnamefont {T.~E.}\ \bibnamefont {Mehlst\"aubler}},\ and\ \bibinfo {author} {\bibfnamefont {K.}~\bibnamefont {Hammerer}},\ }\bibfield  {title} {\bibinfo {title} {Sideband thermometry of ion crystals},\ }\href {https://doi.org/10.1103/PRXQuantum.4.040346} {\bibfield  {journal} {\bibinfo  {journal} {PRX Quantum}\ }\textbf {\bibinfo {volume} {4}},\ \bibinfo {pages} {040346} (\bibinfo {year} {2023})}\BibitemShut {NoStop}%
\bibitem [{\citenamefont {Wesenberg}\ \emph {et~al.}(2007)\citenamefont {Wesenberg}, \citenamefont {Epstein}, \citenamefont {Leibfried}, \citenamefont {Blakestad}, \citenamefont {Britton}, \citenamefont {Home}, \citenamefont {Itano}, \citenamefont {Jost}, \citenamefont {Knill}, \citenamefont {Langer}, \citenamefont {Ozeri}, \citenamefont {Seidelin},\ and\ \citenamefont {Wineland}}]{Wesenberg2007Fluorescence}%
  \BibitemOpen
  \bibfield  {author} {\bibinfo {author} {\bibfnamefont {J.~H.}\ \bibnamefont {Wesenberg}}, \bibinfo {author} {\bibfnamefont {R.~J.}\ \bibnamefont {Epstein}}, \bibinfo {author} {\bibfnamefont {D.}~\bibnamefont {Leibfried}}, \bibinfo {author} {\bibfnamefont {R.~B.}\ \bibnamefont {Blakestad}}, \bibinfo {author} {\bibfnamefont {J.}~\bibnamefont {Britton}}, \bibinfo {author} {\bibfnamefont {J.~P.}\ \bibnamefont {Home}}, \bibinfo {author} {\bibfnamefont {W.~M.}\ \bibnamefont {Itano}}, \bibinfo {author} {\bibfnamefont {J.~D.}\ \bibnamefont {Jost}}, \bibinfo {author} {\bibfnamefont {E.}~\bibnamefont {Knill}}, \bibinfo {author} {\bibfnamefont {C.}~\bibnamefont {Langer}}, \bibinfo {author} {\bibfnamefont {R.}~\bibnamefont {Ozeri}}, \bibinfo {author} {\bibfnamefont {S.}~\bibnamefont {Seidelin}},\ and\ \bibinfo {author} {\bibfnamefont {D.~J.}\ \bibnamefont {Wineland}},\ }\bibfield  {title} {\bibinfo {title} {{Fluorescence during Doppler cooling of a single trapped atom}},\ }\href
  {https://doi.org/10.1103/physreva.76.053416} {\bibfield  {journal} {\bibinfo  {journal} {Phys. Rev. A}\ }\textbf {\bibinfo {volume} {76}},\ \bibinfo {pages} {053416} (\bibinfo {year} {2007})}\BibitemShut {NoStop}%
\bibitem [{\citenamefont {Epstein}\ \emph {et~al.}(2007)\citenamefont {Epstein}, \citenamefont {Seidelin}, \citenamefont {Leibfried}, \citenamefont {Wesenberg}, \citenamefont {Bollinger}, \citenamefont {Amini}, \citenamefont {Blakestad}, \citenamefont {Britton}, \citenamefont {Home}, \citenamefont {Itano}, \citenamefont {Jost}, \citenamefont {Knill}, \citenamefont {Langer}, \citenamefont {Ozeri}, \citenamefont {Shiga},\ and\ \citenamefont {Wineland}}]{Epstein2007Simplified}%
  \BibitemOpen
  \bibfield  {author} {\bibinfo {author} {\bibfnamefont {R.~J.}\ \bibnamefont {Epstein}}, \bibinfo {author} {\bibfnamefont {S.}~\bibnamefont {Seidelin}}, \bibinfo {author} {\bibfnamefont {D.}~\bibnamefont {Leibfried}}, \bibinfo {author} {\bibfnamefont {J.~H.}\ \bibnamefont {Wesenberg}}, \bibinfo {author} {\bibfnamefont {J.~J.}\ \bibnamefont {Bollinger}}, \bibinfo {author} {\bibfnamefont {J.~M.}\ \bibnamefont {Amini}}, \bibinfo {author} {\bibfnamefont {R.~B.}\ \bibnamefont {Blakestad}}, \bibinfo {author} {\bibfnamefont {J.}~\bibnamefont {Britton}}, \bibinfo {author} {\bibfnamefont {J.~P.}\ \bibnamefont {Home}}, \bibinfo {author} {\bibfnamefont {W.~M.}\ \bibnamefont {Itano}}, \bibinfo {author} {\bibfnamefont {J.~D.}\ \bibnamefont {Jost}}, \bibinfo {author} {\bibfnamefont {E.}~\bibnamefont {Knill}}, \bibinfo {author} {\bibfnamefont {C.}~\bibnamefont {Langer}}, \bibinfo {author} {\bibfnamefont {R.}~\bibnamefont {Ozeri}}, \bibinfo {author} {\bibfnamefont {N.}~\bibnamefont {Shiga}},\ and\ \bibinfo {author}
  {\bibfnamefont {D.~J.}\ \bibnamefont {Wineland}},\ }\bibfield  {title} {\bibinfo {title} {Simplified motional heating rate measurements of trapped ions},\ }\href {https://doi.org/10.1103/PhysRevA.76.033411} {\bibfield  {journal} {\bibinfo  {journal} {Phys. Rev. A}\ }\textbf {\bibinfo {volume} {76}},\ \bibinfo {pages} {033411} (\bibinfo {year} {2007})}\BibitemShut {NoStop}%
\bibitem [{\citenamefont {Zipkes}\ \emph {et~al.}(2010)\citenamefont {Zipkes}, \citenamefont {Palzer}, \citenamefont {Sias},\ and\ \citenamefont {Köhl}}]{Zipkes2010A}%
  \BibitemOpen
  \bibfield  {author} {\bibinfo {author} {\bibfnamefont {C.}~\bibnamefont {Zipkes}}, \bibinfo {author} {\bibfnamefont {S.}~\bibnamefont {Palzer}}, \bibinfo {author} {\bibfnamefont {C.}~\bibnamefont {Sias}},\ and\ \bibinfo {author} {\bibfnamefont {M.}~\bibnamefont {Köhl}},\ }\bibfield  {title} {\bibinfo {title} {{A trapped single ion inside a Bose–Einstein condensate}},\ }\href {https://doi.org/10.1038/nature08865} {\bibfield  {journal} {\bibinfo  {journal} {Nature}\ }\textbf {\bibinfo {volume} {464}},\ \bibinfo {pages} {388} (\bibinfo {year} {2010})}\BibitemShut {NoStop}%
\bibitem [{\citenamefont {Sikorsky}\ \emph {et~al.}(2017)\citenamefont {Sikorsky}, \citenamefont {Meir}, \citenamefont {Akerman}, \citenamefont {Ben-shlomi},\ and\ \citenamefont {Ozeri}}]{Sikorsky2017Doppler}%
  \BibitemOpen
  \bibfield  {author} {\bibinfo {author} {\bibfnamefont {T.}~\bibnamefont {Sikorsky}}, \bibinfo {author} {\bibfnamefont {Z.}~\bibnamefont {Meir}}, \bibinfo {author} {\bibfnamefont {N.}~\bibnamefont {Akerman}}, \bibinfo {author} {\bibfnamefont {R.}~\bibnamefont {Ben-shlomi}},\ and\ \bibinfo {author} {\bibfnamefont {R.}~\bibnamefont {Ozeri}},\ }\bibfield  {title} {\bibinfo {title} {{Doppler cooling thermometry of a multilevel ion in the presence of micromotion}},\ }\href {https://doi.org/10.1103/physreva.96.012519} {\bibfield  {journal} {\bibinfo  {journal} {Phys. Rev. A}\ }\textbf {\bibinfo {volume} {96}},\ \bibinfo {pages} {012519} (\bibinfo {year} {2017})}\BibitemShut {NoStop}%
\bibitem [{\citenamefont {Nötzold}\ \emph {et~al.}(2020)\citenamefont {Nötzold}, \citenamefont {Hassan}, \citenamefont {Tauch}, \citenamefont {Endres}, \citenamefont {Wester},\ and\ \citenamefont {Weidemüller}}]{notzold2020thermometry}%
  \BibitemOpen
  \bibfield  {author} {\bibinfo {author} {\bibfnamefont {M.}~\bibnamefont {Nötzold}}, \bibinfo {author} {\bibfnamefont {S.~Z.}\ \bibnamefont {Hassan}}, \bibinfo {author} {\bibfnamefont {J.}~\bibnamefont {Tauch}}, \bibinfo {author} {\bibfnamefont {E.}~\bibnamefont {Endres}}, \bibinfo {author} {\bibfnamefont {R.}~\bibnamefont {Wester}},\ and\ \bibinfo {author} {\bibfnamefont {M.}~\bibnamefont {Weidemüller}},\ }\bibfield  {title} {\bibinfo {title} {Thermometry in a multipole ion trap},\ }\href {https://www.mdpi.com/2076-3417/10/15/5264} {\bibfield  {journal} {\bibinfo  {journal} {Appl. Sci.}\ }\textbf {\bibinfo {volume} {10}} (\bibinfo {year} {2020})}\BibitemShut {NoStop}%
\bibitem [{\citenamefont {Rouse}\ and\ \citenamefont {Willitsch}(2015)}]{rouse15a}%
  \BibitemOpen
  \bibfield  {author} {\bibinfo {author} {\bibfnamefont {I.}~\bibnamefont {Rouse}}\ and\ \bibinfo {author} {\bibfnamefont {S.}~\bibnamefont {Willitsch}},\ }\bibfield  {title} {\bibinfo {title} {Superstatistical velocity distributions of cold trapped ions in molecular-dynamics simulations},\ }\href {https://doi.org/10.1103/PhysRevA.92.053420} {\bibfield  {journal} {\bibinfo  {journal} {Phys. Rev. A}\ }\textbf {\bibinfo {volume} {92}},\ \bibinfo {pages} {053420} (\bibinfo {year} {2015})}\BibitemShut {NoStop}%
\bibitem [{\citenamefont {Mangeng}\ \emph {et~al.}(2023)\citenamefont {Mangeng}, \citenamefont {Yin}, \citenamefont {Karl},\ and\ \citenamefont {Willitsch}}]{mangeng2023}%
  \BibitemOpen
  \bibfield  {author} {\bibinfo {author} {\bibfnamefont {C.}~\bibnamefont {Mangeng}}, \bibinfo {author} {\bibfnamefont {Y.}~\bibnamefont {Yin}}, \bibinfo {author} {\bibfnamefont {R.}~\bibnamefont {Karl}},\ and\ \bibinfo {author} {\bibfnamefont {S.}~\bibnamefont {Willitsch}},\ }\bibfield  {title} {\bibinfo {title} {Experimental implementation of laser cooling of trapped ions in strongly inhomogeneous magnetic fields},\ }\href {https://doi.org/10.1103/PhysRevResearch.5.043180} {\bibfield  {journal} {\bibinfo  {journal} {Phys. Rev. Res.}\ }\textbf {\bibinfo {volume} {5}},\ \bibinfo {pages} {043180} (\bibinfo {year} {2023})}\BibitemShut {NoStop}%
\bibitem [{\citenamefont {Eastman}\ \emph {et~al.}(2017)\citenamefont {Eastman}, \citenamefont {Swails}, \citenamefont {Chodera}, \citenamefont {McGibbon}, \citenamefont {Zhao}, \citenamefont {Beauchamp}, \citenamefont {Wang}, \citenamefont {Simmonett}, \citenamefont {Harrigan}, \citenamefont {Stern}, \citenamefont {Wiewiora}, \citenamefont {Brooks},\ and\ \citenamefont {Pande}}]{openmm}%
  \BibitemOpen
  \bibfield  {author} {\bibinfo {author} {\bibfnamefont {P.}~\bibnamefont {Eastman}}, \bibinfo {author} {\bibfnamefont {J.}~\bibnamefont {Swails}}, \bibinfo {author} {\bibfnamefont {J.~D.}\ \bibnamefont {Chodera}}, \bibinfo {author} {\bibfnamefont {R.~T.}\ \bibnamefont {McGibbon}}, \bibinfo {author} {\bibfnamefont {Y.}~\bibnamefont {Zhao}}, \bibinfo {author} {\bibfnamefont {K.~A.}\ \bibnamefont {Beauchamp}}, \bibinfo {author} {\bibfnamefont {L.-P.}\ \bibnamefont {Wang}}, \bibinfo {author} {\bibfnamefont {A.~C.}\ \bibnamefont {Simmonett}}, \bibinfo {author} {\bibfnamefont {M.~P.}\ \bibnamefont {Harrigan}}, \bibinfo {author} {\bibfnamefont {C.~D.}\ \bibnamefont {Stern}}, \bibinfo {author} {\bibfnamefont {R.~P.}\ \bibnamefont {Wiewiora}}, \bibinfo {author} {\bibfnamefont {B.~R.}\ \bibnamefont {Brooks}},\ and\ \bibinfo {author} {\bibfnamefont {V.~S.}\ \bibnamefont {Pande}},\ }\bibfield  {title} {\bibinfo {title} {Openmm 7: Rapid development of high performance algorithms for molecular dynamics},\ }\href
  {https://doi.org/10.1371/journal.pcbi.1005659} {\bibfield  {journal} {\bibinfo  {journal} {PLoS Comput. Biol.}\ }\textbf {\bibinfo {volume} {13}},\ \bibinfo {pages} {1} (\bibinfo {year} {2017})}\BibitemShut {NoStop}%
\bibitem [{\citenamefont {LeCun}\ \emph {et~al.}(2015)\citenamefont {LeCun}, \citenamefont {Bengio},\ and\ \citenamefont {Hinton}}]{lecun2015deep}%
  \BibitemOpen
  \bibfield  {author} {\bibinfo {author} {\bibfnamefont {Y.}~\bibnamefont {LeCun}}, \bibinfo {author} {\bibfnamefont {Y.}~\bibnamefont {Bengio}},\ and\ \bibinfo {author} {\bibfnamefont {G.}~\bibnamefont {Hinton}},\ }\bibfield  {title} {\bibinfo {title} {Deep learning},\ }\href {https://doi.org/https://doi.org/10.1038/nature14539} {\bibfield  {journal} {\bibinfo  {journal} {nature}\ }\textbf {\bibinfo {volume} {521}},\ \bibinfo {pages} {436} (\bibinfo {year} {2015})}\BibitemShut {NoStop}%
\bibitem [{\citenamefont {Krizhevsky}\ \emph {et~al.}(2017)\citenamefont {Krizhevsky}, \citenamefont {Sutskever},\ and\ \citenamefont {Hinton}}]{krizhevsky2012imagenet}%
  \BibitemOpen
  \bibfield  {author} {\bibinfo {author} {\bibfnamefont {A.}~\bibnamefont {Krizhevsky}}, \bibinfo {author} {\bibfnamefont {I.}~\bibnamefont {Sutskever}},\ and\ \bibinfo {author} {\bibfnamefont {G.~E.}\ \bibnamefont {Hinton}},\ }\bibfield  {title} {\bibinfo {title} {Imagenet classification with deep convolutional neural networks},\ }\href {https://doi.org/10.1145/3065386} {\bibfield  {journal} {\bibinfo  {journal} {Commun. ACM}\ }\textbf {\bibinfo {volume} {60}},\ \bibinfo {pages} {84} (\bibinfo {year} {2017})}\BibitemShut {NoStop}%
\bibitem [{\citenamefont {Simonyan}\ and\ \citenamefont {Zisserman}(2015)}]{simonyan2014very}%
  \BibitemOpen
  \bibfield  {author} {\bibinfo {author} {\bibfnamefont {K.}~\bibnamefont {Simonyan}}\ and\ \bibinfo {author} {\bibfnamefont {A.}~\bibnamefont {Zisserman}},\ }\href {https://arxiv.org/abs/1409.1556} {\bibinfo {title} {Very deep convolutional networks for large-scale image recognition}} (\bibinfo {year} {2015}),\ \Eprint {https://arxiv.org/abs/1409.1556} {arXiv:1409.1556 [cs.CV]} \BibitemShut {NoStop}%
\bibitem [{\citenamefont {He}\ \emph {et~al.}(2016)\citenamefont {He}, \citenamefont {Zhang}, \citenamefont {Ren},\ and\ \citenamefont {Sun}}]{he2016deep}%
  \BibitemOpen
  \bibfield  {author} {\bibinfo {author} {\bibfnamefont {K.}~\bibnamefont {He}}, \bibinfo {author} {\bibfnamefont {X.}~\bibnamefont {Zhang}}, \bibinfo {author} {\bibfnamefont {S.}~\bibnamefont {Ren}},\ and\ \bibinfo {author} {\bibfnamefont {J.}~\bibnamefont {Sun}},\ }\bibfield  {title} {\bibinfo {title} {Deep residual learning for image recognition},\ }in\ \href {https://openaccess.thecvf.com/content_cvpr_2016/html/He_Deep_Residual_Learning_CVPR_2016_paper.html} {\emph {\bibinfo {booktitle} {Proceedings of the IEEE Conference on Computer Vision and Pattern Recognition}}},\ \bibinfo {address} {Las Vegas, 2016}\ (\bibinfo  {publisher} {IEEE},\ \bibinfo {address} {New York},\ \bibinfo {year} {2016})\ pp.\ \bibinfo {pages} {770--778}\BibitemShut {NoStop}%
\bibitem [{\citenamefont {Tan}\ \emph {et~al.}(2019)\citenamefont {Tan}, \citenamefont {Chen}, \citenamefont {Pang}, \citenamefont {Vasudevan}, \citenamefont {Sandler}, \citenamefont {Howard},\ and\ \citenamefont {Le}}]{tan2019mnasnet}%
  \BibitemOpen
  \bibfield  {author} {\bibinfo {author} {\bibfnamefont {M.}~\bibnamefont {Tan}}, \bibinfo {author} {\bibfnamefont {B.}~\bibnamefont {Chen}}, \bibinfo {author} {\bibfnamefont {R.}~\bibnamefont {Pang}}, \bibinfo {author} {\bibfnamefont {V.}~\bibnamefont {Vasudevan}}, \bibinfo {author} {\bibfnamefont {M.}~\bibnamefont {Sandler}}, \bibinfo {author} {\bibfnamefont {A.}~\bibnamefont {Howard}},\ and\ \bibinfo {author} {\bibfnamefont {Q.~V.}\ \bibnamefont {Le}},\ }\bibfield  {title} {\bibinfo {title} {Mnasnet: Platform-aware neural architecture search for mobile},\ }in\ \href {https://openaccess.thecvf.com/content_CVPR_2019/html/Tan_MnasNet_Platform-Aware_Neural_Architecture_Search_for_Mobile_CVPR_2019_paper} {\emph {\bibinfo {booktitle} {Proceedings of the IEEE/CVF Conference on Computer Vision and Pattern Recognition}}}\ (\bibinfo  {publisher} {IEEE},\ \bibinfo {address} {New York},\ \bibinfo {year} {2019})\ pp.\ \bibinfo {pages} {2820--2828}\BibitemShut {NoStop}%
\bibitem [{\citenamefont {Yosinski}\ \emph {et~al.}(2014)\citenamefont {Yosinski}, \citenamefont {Clune}, \citenamefont {Bengio},\ and\ \citenamefont {Lipson}}]{yosinski2014transferable}%
  \BibitemOpen
  \bibfield  {author} {\bibinfo {author} {\bibfnamefont {J.}~\bibnamefont {Yosinski}}, \bibinfo {author} {\bibfnamefont {J.}~\bibnamefont {Clune}}, \bibinfo {author} {\bibfnamefont {Y.}~\bibnamefont {Bengio}},\ and\ \bibinfo {author} {\bibfnamefont {H.}~\bibnamefont {Lipson}},\ }\bibfield  {title} {\bibinfo {title} {How transferable are features in deep neural networks?},\ }in\ \href {https://proceedings.neurips.cc/paper_files/paper/2014/file/532a2f85b6977104bc93f8580abbb330-Paper.pdf} {\emph {\bibinfo {booktitle} {Advances in Neural Information Processing Systems 27}}}\ (\bibinfo  {publisher} {Curran Associates, Inc.},\ \bibinfo {address} {Red Hook, NY, USA},\ \bibinfo {year} {2014})\ pp.\ \bibinfo {pages} {3320--3328}\BibitemShut {NoStop}%
\bibitem [{\citenamefont {Oquab}\ \emph {et~al.}(2014)\citenamefont {Oquab}, \citenamefont {Bottou}, \citenamefont {Laptev},\ and\ \citenamefont {Sivic}}]{oquab2014learning}%
  \BibitemOpen
  \bibfield  {author} {\bibinfo {author} {\bibfnamefont {M.}~\bibnamefont {Oquab}}, \bibinfo {author} {\bibfnamefont {L.}~\bibnamefont {Bottou}}, \bibinfo {author} {\bibfnamefont {I.}~\bibnamefont {Laptev}},\ and\ \bibinfo {author} {\bibfnamefont {J.}~\bibnamefont {Sivic}},\ }\bibfield  {title} {\bibinfo {title} {Learning and transferring mid-level image representations using convolutional neural networks},\ }in\ \href {https://openaccess.thecvf.com/content_cvpr_2014/html/Oquab_Learning_and_Transferring_2014_CVPR_paper.html} {\emph {\bibinfo {booktitle} {Proceedings of the IEEE Conference on Computer Vision and Pattern Recognition}}}\ (\bibinfo  {publisher} {IEEE},\ \bibinfo {address} {New York},\ \bibinfo {year} {2014})\ pp.\ \bibinfo {pages} {1717--1724}\BibitemShut {NoStop}%
\bibitem [{\citenamefont {Donahue}\ \emph {et~al.}(2014)\citenamefont {Donahue}, \citenamefont {Jia}, \citenamefont {Vinyals}, \citenamefont {Hoffman}, \citenamefont {Zhang}, \citenamefont {Tzeng},\ and\ \citenamefont {Darrell}}]{donahue2014decaf}%
  \BibitemOpen
  \bibfield  {author} {\bibinfo {author} {\bibfnamefont {J.}~\bibnamefont {Donahue}}, \bibinfo {author} {\bibfnamefont {Y.}~\bibnamefont {Jia}}, \bibinfo {author} {\bibfnamefont {O.}~\bibnamefont {Vinyals}}, \bibinfo {author} {\bibfnamefont {J.}~\bibnamefont {Hoffman}}, \bibinfo {author} {\bibfnamefont {N.}~\bibnamefont {Zhang}}, \bibinfo {author} {\bibfnamefont {E.}~\bibnamefont {Tzeng}},\ and\ \bibinfo {author} {\bibfnamefont {T.}~\bibnamefont {Darrell}},\ }\bibfield  {title} {\bibinfo {title} {Decaf: A deep convolutional activation feature for generic visual recognition},\ }in\ \href {https://proceedings.mlr.press/v32/donahue14.html} {\emph {\bibinfo {booktitle} {Proceedings of the 31st International Conference on Machine Learning}}},\ \bibinfo {series} {Proceedings of Machine Learning Research}, Vol.~\bibinfo {volume} {32},\ \bibinfo {editor} {edited by\ \bibinfo {editor} {\bibfnamefont {E.~P.}\ \bibnamefont {Xing}}\ and\ \bibinfo {editor} {\bibfnamefont {T.}~\bibnamefont {Jebara}}}\ (\bibinfo
  {publisher} {PMLR},\ \bibinfo {address} {Bejing, China},\ \bibinfo {year} {2014})\ pp.\ \bibinfo {pages} {647--655}\BibitemShut {NoStop}%
\bibitem [{\citenamefont {Paszke}\ \emph {et~al.}(2019)\citenamefont {Paszke}, \citenamefont {Gross}, \citenamefont {Massa}, \citenamefont {Lerer}, \citenamefont {Bradbury}, \citenamefont {Chanan}, \citenamefont {Killeen}, \citenamefont {Lin}, \citenamefont {Gimelshein}, \citenamefont {Antiga}, \citenamefont {Desmaison}, \citenamefont {Kopf}, \citenamefont {Yang}, \citenamefont {DeVito}, \citenamefont {Raison}, \citenamefont {Tejani}, \citenamefont {Chilamkurthy}, \citenamefont {Steiner}, \citenamefont {Fang}, \citenamefont {Bai},\ and\ \citenamefont {Chintala}}]{paszke2019pytorch}%
  \BibitemOpen
  \bibfield  {author} {\bibinfo {author} {\bibfnamefont {A.}~\bibnamefont {Paszke}}, \bibinfo {author} {\bibfnamefont {S.}~\bibnamefont {Gross}}, \bibinfo {author} {\bibfnamefont {F.}~\bibnamefont {Massa}}, \bibinfo {author} {\bibfnamefont {A.}~\bibnamefont {Lerer}}, \bibinfo {author} {\bibfnamefont {J.}~\bibnamefont {Bradbury}}, \bibinfo {author} {\bibfnamefont {G.}~\bibnamefont {Chanan}}, \bibinfo {author} {\bibfnamefont {T.}~\bibnamefont {Killeen}}, \bibinfo {author} {\bibfnamefont {Z.}~\bibnamefont {Lin}}, \bibinfo {author} {\bibfnamefont {N.}~\bibnamefont {Gimelshein}}, \bibinfo {author} {\bibfnamefont {L.}~\bibnamefont {Antiga}}, \bibinfo {author} {\bibfnamefont {A.}~\bibnamefont {Desmaison}}, \bibinfo {author} {\bibfnamefont {A.}~\bibnamefont {Kopf}}, \bibinfo {author} {\bibfnamefont {E.}~\bibnamefont {Yang}}, \bibinfo {author} {\bibfnamefont {Z.}~\bibnamefont {DeVito}}, \bibinfo {author} {\bibfnamefont {M.}~\bibnamefont {Raison}}, \bibinfo {author} {\bibfnamefont {A.}~\bibnamefont {Tejani}}, \bibinfo
  {author} {\bibfnamefont {S.}~\bibnamefont {Chilamkurthy}}, \bibinfo {author} {\bibfnamefont {B.}~\bibnamefont {Steiner}}, \bibinfo {author} {\bibfnamefont {L.}~\bibnamefont {Fang}}, \bibinfo {author} {\bibfnamefont {J.}~\bibnamefont {Bai}},\ and\ \bibinfo {author} {\bibfnamefont {S.}~\bibnamefont {Chintala}},\ }\bibfield  {title} {\bibinfo {title} {Pytorch: An imperative style, high-performance deep learning library},\ }in\ \href {http://papers.neurips.cc/paper/9015-pytorch-an-imperative-style-high-performance-deep-learning-library.pdf} {\emph {\bibinfo {booktitle} {Advances in Neural Information Processing Systems 32}}}\ (\bibinfo  {publisher} {Curran Associates, Inc.},\ \bibinfo {address} {Red Hook, NY, USA},\ \bibinfo {year} {2019})\ pp.\ \bibinfo {pages} {8024--8035}\BibitemShut {NoStop}%
\bibitem [{\citenamefont {Deng}\ \emph {et~al.}(2009)\citenamefont {Deng}, \citenamefont {Dong}, \citenamefont {Socher}, \citenamefont {Li}, \citenamefont {Li},\ and\ \citenamefont {Fei-Fei}}]{deng2009imagenet}%
  \BibitemOpen
  \bibfield  {author} {\bibinfo {author} {\bibfnamefont {J.}~\bibnamefont {Deng}}, \bibinfo {author} {\bibfnamefont {W.}~\bibnamefont {Dong}}, \bibinfo {author} {\bibfnamefont {R.}~\bibnamefont {Socher}}, \bibinfo {author} {\bibfnamefont {L.-J.}\ \bibnamefont {Li}}, \bibinfo {author} {\bibfnamefont {K.}~\bibnamefont {Li}},\ and\ \bibinfo {author} {\bibfnamefont {L.}~\bibnamefont {Fei-Fei}},\ }\bibfield  {title} {\bibinfo {title} {Imagenet: A large-scale hierarchical image database},\ }in\ \href {https://doi.org/10.1109/CVPR.2009.5206848} {\emph {\bibinfo {booktitle} {2009 IEEE Conference on Computer Vision and Pattern Recognition}}}\ (\bibinfo  {publisher} {IEEE},\ \bibinfo {address} {New York},\ \bibinfo {year} {2009})\ pp.\ \bibinfo {pages} {248--255}\BibitemShut {NoStop}%
\bibitem [{\citenamefont {Clark}\ and\ \citenamefont {Contributors}(2024)}]{pillow}%
  \BibitemOpen
  \bibfield  {author} {\bibinfo {author} {\bibfnamefont {J.~A.}\ \bibnamefont {Clark}}\ and\ \bibinfo {author} {\bibnamefont {Contributors}},\ }\href@noop {} {\bibinfo {title} {Pillow (pil fork)}},\ \bibinfo {howpublished} {\url{https://python-pillow.github.io/}} (\bibinfo {year} {2024}),\ \bibinfo {note} {version 10.3.0}\BibitemShut {NoStop}%
\bibitem [{\citenamefont {Goodfellow}\ \emph {et~al.}(2016)\citenamefont {Goodfellow}, \citenamefont {Bengio},\ and\ \citenamefont {Courville}}]{Goodfellow-et-al-2016}%
  \BibitemOpen
  \bibfield  {author} {\bibinfo {author} {\bibfnamefont {I.}~\bibnamefont {Goodfellow}}, \bibinfo {author} {\bibfnamefont {Y.}~\bibnamefont {Bengio}},\ and\ \bibinfo {author} {\bibfnamefont {A.}~\bibnamefont {Courville}},\ }\href@noop {} {\emph {\bibinfo {title} {Deep Learning}}}\ (\bibinfo  {publisher} {MIT Press, Cambridge, MA},\ \bibinfo {year} {2016})\ \bibinfo {note} {\url{http://www.deeplearningbook.org}}\BibitemShut {NoStop}%
\bibitem [{\citenamefont {Caruana}\ \emph {et~al.}(2000)\citenamefont {Caruana}, \citenamefont {Lawrence},\ and\ \citenamefont {Giles}}]{caruana2000overfitting}%
  \BibitemOpen
  \bibfield  {author} {\bibinfo {author} {\bibfnamefont {R.}~\bibnamefont {Caruana}}, \bibinfo {author} {\bibfnamefont {S.}~\bibnamefont {Lawrence}},\ and\ \bibinfo {author} {\bibfnamefont {C.}~\bibnamefont {Giles}},\ }\bibfield  {title} {\bibinfo {title} {Overfitting in neural nets: Backpropagation, conjugate gradient, and early stopping},\ }in\ \href {https://proceedings.neurips.cc/paper_files/paper/2000/file/059fdcd96baeb75112f09fa1dcc740cc-Paper.pdf} {\emph {\bibinfo {booktitle} {Advances in Neural Information Processing Systems 13}}}\ (\bibinfo  {publisher} {MIT Press},\ \bibinfo {address} {Cambridge, MA, USA},\ \bibinfo {year} {2000})\ p.\ \bibinfo {pages} {381–387}\BibitemShut {NoStop}%
\bibitem [{\citenamefont {Prechelt}(1998)}]{prechelt2002early}%
  \BibitemOpen
  \bibfield  {author} {\bibinfo {author} {\bibfnamefont {L.}~\bibnamefont {Prechelt}},\ }\bibfield  {title} {\bibinfo {title} {Early stopping - but when?},\ }in\ \href {https://doi.org/10.1007/3-540-49430-8_3} {\emph {\bibinfo {booktitle} {Neural Networks: Tricks of the Trade}}},\ \bibinfo {editor} {edited by\ \bibinfo {editor} {\bibfnamefont {G.~B.}\ \bibnamefont {Orr}}\ and\ \bibinfo {editor} {\bibfnamefont {K.-R.}\ \bibnamefont {M{\"u}ller}}}\ (\bibinfo  {publisher} {Springer Berlin Heidelberg},\ \bibinfo {year} {1998})\ pp.\ \bibinfo {pages} {55--69}\BibitemShut {NoStop}%
\bibitem [{\citenamefont {Hawkins}(2004)}]{hawkins2004problem}%
  \BibitemOpen
  \bibfield  {author} {\bibinfo {author} {\bibfnamefont {D.~M.}\ \bibnamefont {Hawkins}},\ }\bibfield  {title} {\bibinfo {title} {The problem of overfitting},\ }\href {https://doi.org/10.1021/ci0342472} {\bibfield  {journal} {\bibinfo  {journal} {J. Chem. Inf. Comput. Sci.}\ }\textbf {\bibinfo {volume} {44}},\ \bibinfo {pages} {1} (\bibinfo {year} {2004})},\ \bibinfo {note} {pMID: 14741005}\BibitemShut {NoStop}%
\bibitem [{\citenamefont {Yao}\ \emph {et~al.}(2007)\citenamefont {Yao}, \citenamefont {Rosasco},\ and\ \citenamefont {Caponnetto}}]{yao2007early}%
  \BibitemOpen
  \bibfield  {author} {\bibinfo {author} {\bibfnamefont {Y.}~\bibnamefont {Yao}}, \bibinfo {author} {\bibfnamefont {L.}~\bibnamefont {Rosasco}},\ and\ \bibinfo {author} {\bibfnamefont {A.}~\bibnamefont {Caponnetto}},\ }\bibfield  {title} {\bibinfo {title} {On early stopping in gradient descent learning},\ }\href {https://doi.org/https://doi.org/10.1007/s00365-006-0663-2} {\bibfield  {journal} {\bibinfo  {journal} {Constr. Approx.}\ }\textbf {\bibinfo {volume} {26}},\ \bibinfo {pages} {289} (\bibinfo {year} {2007})}\BibitemShut {NoStop}%
\bibitem [{\citenamefont {Perez}\ and\ \citenamefont {Wang}(2017)}]{perez2017effectiveness}%
  \BibitemOpen
  \bibfield  {author} {\bibinfo {author} {\bibfnamefont {L.}~\bibnamefont {Perez}}\ and\ \bibinfo {author} {\bibfnamefont {J.}~\bibnamefont {Wang}},\ }\href {https://arxiv.org/abs/1712.04621} {\bibinfo {title} {The effectiveness of data augmentation in image classification using deep learning}} (\bibinfo {year} {2017}),\ \Eprint {https://arxiv.org/abs/1712.04621} {arXiv:1712.04621 [cs.CV]} \BibitemShut {NoStop}%
\bibitem [{\citenamefont {Shorten}\ and\ \citenamefont {Khoshgoftaar}(2019)}]{shorten2019survey}%
  \BibitemOpen
  \bibfield  {author} {\bibinfo {author} {\bibfnamefont {C.}~\bibnamefont {Shorten}}\ and\ \bibinfo {author} {\bibfnamefont {T.~M.}\ \bibnamefont {Khoshgoftaar}},\ }\bibfield  {title} {\bibinfo {title} {A survey on image data augmentation for deep learning},\ }\href {https://doi.org/10.1186/s40537-019-0197-0} {\bibfield  {journal} {\bibinfo  {journal} {J. Big Data}\ }\textbf {\bibinfo {volume} {6}},\ \bibinfo {pages} {1} (\bibinfo {year} {2019})}\BibitemShut {NoStop}%
\bibitem [{\citenamefont {Allsopp}\ \emph {et~al.}(2025)\citenamefont {Allsopp}, \citenamefont {Diprose}, \citenamefont {Heazlewood}, \citenamefont {Zagorec-Marks}, \citenamefont {Lewandowski}, \citenamefont {Petralia},\ and\ \citenamefont {Softley}}]{allsopp2025conv}%
  \BibitemOpen
  \bibfield  {author} {\bibinfo {author} {\bibfnamefont {J.}~\bibnamefont {Allsopp}}, \bibinfo {author} {\bibfnamefont {J.}~\bibnamefont {Diprose}}, \bibinfo {author} {\bibfnamefont {B.~R.}\ \bibnamefont {Heazlewood}}, \bibinfo {author} {\bibfnamefont {C.}~\bibnamefont {Zagorec-Marks}}, \bibinfo {author} {\bibfnamefont {H.~J.}\ \bibnamefont {Lewandowski}}, \bibinfo {author} {\bibfnamefont {L.~S.}\ \bibnamefont {Petralia}},\ and\ \bibinfo {author} {\bibfnamefont {T.~P.}\ \bibnamefont {Softley}},\ }\bibfield  {title} {\bibinfo {title} {Convolutional neural network approach to ion coulomb crystal image analysis},\ }\href {https://doi.org/10.1063/5.0272967} {\bibfield  {journal} {\bibinfo  {journal} {J. Chem. Phys.}\ }\textbf {\bibinfo {volume} {163}},\ \bibinfo {pages} {044201} (\bibinfo {year} {2025})}\BibitemShut {NoStop}%
\bibitem [{\citenamefont {Hudson}(2016)}]{hudson2016sympathetic}%
  \BibitemOpen
  \bibfield  {author} {\bibinfo {author} {\bibfnamefont {E.~R.}\ \bibnamefont {Hudson}},\ }\bibfield  {title} {\bibinfo {title} {Sympathetic cooling of molecular ions with ultracold atoms},\ }\href {https://doi.org/https://doi.org/10.1140/epjti/s40485-016-0035-0} {\bibfield  {journal} {\bibinfo  {journal} {EPJ Techn. Instrum.}\ }\textbf {\bibinfo {volume} {3}},\ \bibinfo {pages} {1} (\bibinfo {year} {2016})}\BibitemShut {NoStop}%
\bibitem [{\citenamefont {Tomza}\ \emph {et~al.}(2019)\citenamefont {Tomza}, \citenamefont {Jachymski}, \citenamefont {Gerritsma}, \citenamefont {Negretti}, \citenamefont {Calarco}, \citenamefont {Idziaszek},\ and\ \citenamefont {Julienne}}]{Tomza2019Cold}%
  \BibitemOpen
  \bibfield  {author} {\bibinfo {author} {\bibfnamefont {M.}~\bibnamefont {Tomza}}, \bibinfo {author} {\bibfnamefont {K.}~\bibnamefont {Jachymski}}, \bibinfo {author} {\bibfnamefont {R.}~\bibnamefont {Gerritsma}}, \bibinfo {author} {\bibfnamefont {A.}~\bibnamefont {Negretti}}, \bibinfo {author} {\bibfnamefont {T.}~\bibnamefont {Calarco}}, \bibinfo {author} {\bibfnamefont {Z.}~\bibnamefont {Idziaszek}},\ and\ \bibinfo {author} {\bibfnamefont {P.~S.}\ \bibnamefont {Julienne}},\ }\bibfield  {title} {\bibinfo {title} {Cold hybrid ion-atom systems},\ }\href {https://doi.org/10.1103/RevModPhys.91.035001} {\bibfield  {journal} {\bibinfo  {journal} {Rev. Mod. Phys.}\ }\textbf {\bibinfo {volume} {91}},\ \bibinfo {pages} {035001} (\bibinfo {year} {2019})}\BibitemShut {NoStop}%
\bibitem [{\citenamefont {Willitsch}(2015)}]{willitsch2015ion}%
  \BibitemOpen
  \bibfield  {author} {\bibinfo {author} {\bibfnamefont {S.}~\bibnamefont {Willitsch}},\ }\bibfield  {title} {\bibinfo {title} {Ion-atom hybrid systems},\ }\href {https://doi.org/10.3254/978-1-61499-526-5-255} {\bibfield  {journal} {\bibinfo  {journal} {Proc. Int. Sch. Phys. Enrico Fermi}\ }\textbf {\bibinfo {volume} {189}},\ \bibinfo {pages} {255} (\bibinfo {year} {2015})}\BibitemShut {NoStop}%
\bibitem [{\citenamefont {Eberle}\ \emph {et~al.}(2015)\citenamefont {Eberle}, \citenamefont {Dörfler}, \citenamefont {von Planta}, \citenamefont {Ravi}, \citenamefont {Haas}, \citenamefont {Zhang}, \citenamefont {van~de Meerakker},\ and\ \citenamefont {Willitsch}}]{eberle2015ion}%
  \BibitemOpen
  \bibfield  {author} {\bibinfo {author} {\bibfnamefont {P.}~\bibnamefont {Eberle}}, \bibinfo {author} {\bibfnamefont {A.~D.}\ \bibnamefont {Dörfler}}, \bibinfo {author} {\bibfnamefont {C.}~\bibnamefont {von Planta}}, \bibinfo {author} {\bibfnamefont {K.}~\bibnamefont {Ravi}}, \bibinfo {author} {\bibfnamefont {D.}~\bibnamefont {Haas}}, \bibinfo {author} {\bibfnamefont {D.}~\bibnamefont {Zhang}}, \bibinfo {author} {\bibfnamefont {S.~Y.~T.}\ \bibnamefont {van~de Meerakker}},\ and\ \bibinfo {author} {\bibfnamefont {S.}~\bibnamefont {Willitsch}},\ }\bibfield  {title} {\bibinfo {title} {Ion-atom and ion-molecule hybrid systems: Ion-neutral chemistry at ultralow energies},\ }\href {https://doi.org/10.1088/1742-6596/635/1/012012} {\bibfield  {journal} {\bibinfo  {journal} {J. Phys.: Conf. Ser.}\ }\textbf {\bibinfo {volume} {635}},\ \bibinfo {pages} {012012} (\bibinfo {year} {2015})}\BibitemShut {NoStop}%
\bibitem [{\citenamefont {D{\"o}rfler}\ \emph {et~al.}(2019)\citenamefont {D{\"o}rfler}, \citenamefont {Eberle}, \citenamefont {Koner}, \citenamefont {Tomza}, \citenamefont {Meuwly},\ and\ \citenamefont {Willitsch}}]{dorfler2019long}%
  \BibitemOpen
  \bibfield  {author} {\bibinfo {author} {\bibfnamefont {A.~D.}\ \bibnamefont {D{\"o}rfler}}, \bibinfo {author} {\bibfnamefont {P.}~\bibnamefont {Eberle}}, \bibinfo {author} {\bibfnamefont {D.}~\bibnamefont {Koner}}, \bibinfo {author} {\bibfnamefont {M.}~\bibnamefont {Tomza}}, \bibinfo {author} {\bibfnamefont {M.}~\bibnamefont {Meuwly}},\ and\ \bibinfo {author} {\bibfnamefont {S.}~\bibnamefont {Willitsch}},\ }\bibfield  {title} {\bibinfo {title} {Long-range versus short-range effects in cold molecular ion-neutral collisions},\ }\href {https://doi.org/https://doi.org/10.1038/s41467-019-13218-x} {\bibfield  {journal} {\bibinfo  {journal} {Nat. Commun.}\ }\textbf {\bibinfo {volume} {10}},\ \bibinfo {pages} {5429} (\bibinfo {year} {2019})}\BibitemShut {NoStop}%
\bibitem [{\citenamefont {Dei{\ss}}\ \emph {et~al.}(2024)\citenamefont {Dei{\ss}}, \citenamefont {Willitsch},\ and\ \citenamefont {Hecker~Denschlag}}]{deiss2024cold}%
  \BibitemOpen
  \bibfield  {author} {\bibinfo {author} {\bibfnamefont {M.}~\bibnamefont {Dei{\ss}}}, \bibinfo {author} {\bibfnamefont {S.}~\bibnamefont {Willitsch}},\ and\ \bibinfo {author} {\bibfnamefont {J.}~\bibnamefont {Hecker~Denschlag}},\ }\bibfield  {title} {\bibinfo {title} {Cold trapped molecular ions and hybrid platforms for ions and neutral particles},\ }\href {https://doi.org/https://doi.org/10.1038/s41567-024-02440-0} {\bibfield  {journal} {\bibinfo  {journal} {Nat. Phys.}\ }\textbf {\bibinfo {volume} {20}},\ \bibinfo {pages} {713–721} (\bibinfo {year} {2024})}\BibitemShut {NoStop}%
\bibitem [{\citenamefont {Yin}\ and\ \citenamefont {Willitsch}(2025)}]{yin2025code}%
  \BibitemOpen
  \bibfield  {author} {\bibinfo {author} {\bibfnamefont {Y.}~\bibnamefont {Yin}}\ and\ \bibinfo {author} {\bibfnamefont {S.}~\bibnamefont {Willitsch}},\ }\href {https://doi.org/10.5281/zenodo.14926184} {\bibinfo {title} {Code for "ion counting and temperature determination of coulomb-crystallized laser-cooled ions in traps using convolutional neural networks"}} (\bibinfo {year} {2025}),\ \bibinfo {note} {dataset}\BibitemShut {NoStop}%
\end{thebibliography}%

\end{document}